\documentclass[manuscript,screen]{acmart}

\usepackage{times}
\usepackage{epsfig}
\usepackage{graphicx}
\usepackage{amsmath}
\usepackage{mathtools}
\usepackage{calligra}
\usepackage{color}
\usepackage{array}
\usepackage{hhline}
\usepackage{subfig}
\usepackage{multirow}
\usepackage{caption}
\usepackage{boxhandler}
\usepackage{tabularx}
\usepackage{adjustbox}
\usepackage{verbatim}
\usepackage{bm}
\usepackage{float}

\AtBeginDocument{%
  \providecommand\BibTeX{{%
    \normalfont B\kern-0.5em{\scshape i\kern-0.25em b}\kern-0.8em\TeX}}}

\setcopyright{acmcopyright}
\acmJournal{TOMM}
\acmYear{2022} \acmVolume{1} \acmNumber{1} \acmArticle{1} \acmMonth{1} \acmPrice{15.00}\acmDOI{10.1145/3532864}

\acmConference[Woodstock '18]{Woodstock '18: ACM Symposium on Neural
  Gaze Detection}{June 03--05, 2018}{Woodstock, NY}
\acmBooktitle{Woodstock '18: ACM Symposium on Neural Gaze Detection,
  June 03--05, 2018, Woodstock, NY}
\acmPrice{15.00}
\acmISBN{978-1-4503-XXXX-X/18/06}

\begin{document}


\title[SHSR]{Sequential Hierarchical Learning with Distribution Transformation for Image Super-Resolution}

\author{Yuqing Liu}
\affiliation{%
	\institution{Dalian University of Technology}
	\country{China}}
\email{liuyuqing@mail.dlut.edu.cn}

\author{Xinfeng Zhang}
\affiliation{%
	\institution{University of the Chinese Academy of Sciences}
	\country{China}}
\email{xfzhang@ucas.ac.cn}

\author{Shanshe Wang}
\affiliation{%
	\institution{Peking University, and Information Technology Research and Development Innovation Center of Peking University, and
	Peng Cheng Laboratory}
	\country{China}}
\email{sswang@pku.edu.cn}

\author{Siwei Ma}
\affiliation{%
	\institution{Peking University, and Information Technology Research and Development Innovation Center of Peking University, and
	Peng Cheng Laboratory}
	\country{China}}
\email{swma@pku.edu.cn}

\author{Wen Gao}
\affiliation{%
	\institution{Peking University, and Information Technology Research and Development Innovation Center of Peking University, and
	Peng Cheng Laboratory}
	\country{China}}
\email{wgao@pku.edu.cn}



\renewcommand{\shortauthors}{Liu, et al.}

\begin{abstract}
    Multi-scale design has been considered in recent image super-resolution (SR) works to explore the hierarchical feature information. 
    Existing multi-scale networks aim to build elaborate blocks or progressive architecture for restoration.
    In general, larger scale features concentrate more on structural and high-level information, while smaller scale features contain plentiful details and textured information. 
    In this point of view, information from larger scale features can be derived from smaller ones.
    Based on the observation, in this paper, we build a sequential hierarchical learning super-resolution network (SHSR) for effective image SR. 
    Specially, we consider the inter-scale correlations of features, and devise a sequential multi-scale block (SMB) to progressively explore the hierarchical information.
    SMB is designed in a recursive way based on the linearity of convolution with restricted parameters.
    Besides the sequential hierarchical learning, we also investigate the correlations among the feature maps and devise a distribution transformation block (DTB).
    Different from attention-based methods, DTB regards the transformation in a normalization manner, and jointly considers the spatial and channel-wise correlations with scaling and bias factors.
    Experiment results show SHSR achieves superior quantitative performance and visual quality to state-of-the-art methods with near 34\% parameters and 50\% MACs off when scaling factor is $\times4$.
    To boost the performance without further training, the extension model SHSR$^+$ with self-ensemble achieves competitive performance than larger networks with near 92\% parameters and 42\% MACs off with scaling factor $\times4$.
\end{abstract}



\maketitle

\section{Introduction}
Single image super-resolution (SISR) aims to recover the high resolution (HR) image from corresponding low resolution (LR) instance. As a highly ill-posed problem, SISR usually suffers from comprehensive degradation situations, such as down-sampling, blurry, and noise. SISR is widely applied in different computer vision tasks, such as facial analysis~\cite{tomm_1}, 3D reconstruction~\cite{tomm_2}, and personal Re-ID~\cite{tomm_3}.

In recent years, convolutional neural networks (CNNs) with elaborate designs have demonstrated impressive restoration performances on SISR problem. SRCNN~\cite{srcnn_pami2016} is the first CNN-based method for SISR, which regards the network as a sparse coding inspired structure. After SRCNN, there are deeper or wider networks with elaborate blocks achieving state-of-the-art performances, such as CARN~\cite{carn_eccv2018}, RDN~\cite{rdn_pami2020}, RFANet~\cite{rfanet_cvpr2020}, and SAN~\cite{san_cvpr2019}. 

Multi-scale design is one of the effective patterns for image restoration and has been considered in advanced SISR works. LapSRN~\cite{lapsrn_cvpr2017}, MS-LapSRN~\cite{lapsrn_pami2019}, and other Laplacian pyramid inspired methods aim to build a progressive restoration network with explicit resolution increase. Utilizing larger feature maps can better illustrate high resolution information with restrict network parameters, but it leads to higher computation complexity when interpolating with larger scaling factors. Another protocol for building a multi-scale network is to explore the features by filters with different receptive fields. MSRN~\cite{msrn_eccv2018}, MDCN~\cite{mdcn_csvt2020}, MGAN~\cite{mgan_csvt2020} and other works consider filters with different receptive fields to explore the hierarchical features. According to the linearity, the larger convolutional layers can be separated into several layers with small kernels and save the parameters.

Multi-scale pattern focuses on the information from different receptive fields, while attention mechanism aims to adjust the distribution of feature maps for effective correlation exploration. Channel-wise attention proposed by Hu~\textit{et al.}~\cite{senet_pami2020} has been investigated in recent SISR works, such as RCAN~\cite{rcan_eccv2018}, IMDN~\cite{imdn_mm2019} and DRN~\cite{drn_cvpr2020}. Channel-wise attention proves to be an efficient block but usually ignores to explore the spatial correlations of features. There are spatial attention networks to concentrate on the pixel-wise correlations. SPARNet~\cite{sparnet_tip2020}, CSFM~\cite{csfm_csvt2020} and other works utilize spatial attention to achieve state-of-the-art performances. However, recent SISR works usually independently consider the transformation from different aspects, and only adjust the features with scaling factors.

\begin{figure}[t]
	\captionsetup[subfloat]{labelformat=empty, justification=centering}
	\begin{center}
		\newcommand{\rowArg}{2.2cm}
		\newcommand{\fullSize}{4.9cm}
		\newcommand{\fullWidth}{7cm}
		\newcommand{\patchSize}{2cm}
		\scriptsize
		\setlength\tabcolsep{0.05cm}
		\begin{tabular}[b]{c c c c c c}
			\multicolumn{3}{c}{\multirow{2}{*}[\rowArg]{
					\subfloat[image\_024 from Urban100~\cite{urban100}]
					{\includegraphics[height=\fullSize, width=\fullWidth]
						{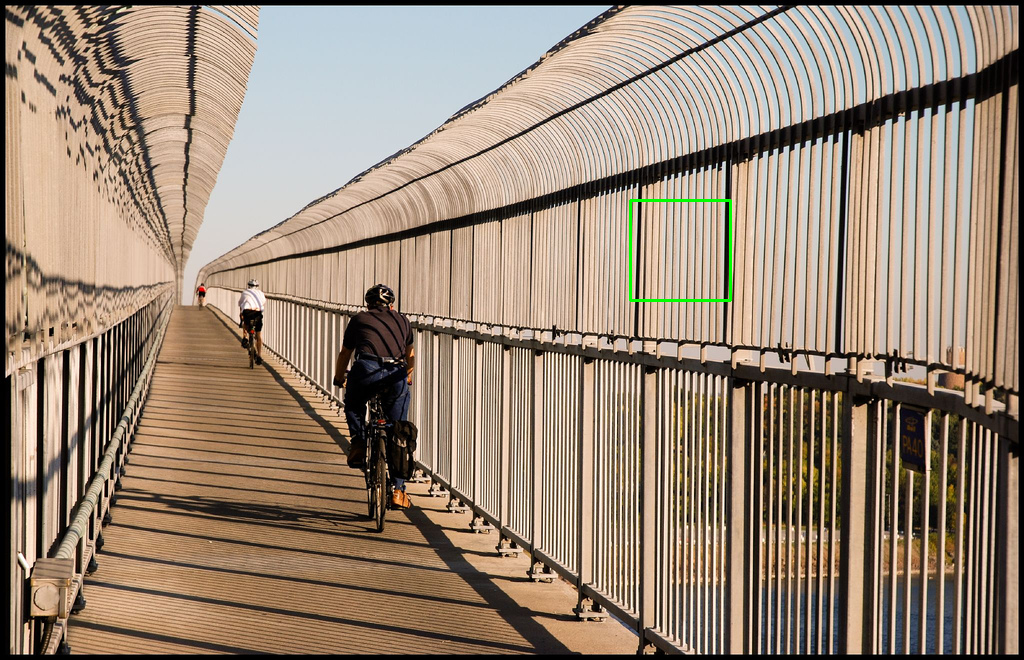}}}} &
			\subfloat[HR~\protect\linebreak(PSNR/SSIM)]
			{\includegraphics[width = \patchSize, height = \patchSize]
				{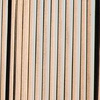}} &
			\subfloat[Bicubic~\protect\linebreak(16.94/0.5539)]
			{\includegraphics[width = \patchSize, height = \patchSize]
				{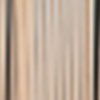}} &
			\subfloat[MSLapSRN~\cite{lapsrn_pami2019} \protect\linebreak(18.10/0.6714)]
			{\includegraphics[width = \patchSize, height = \patchSize]
				{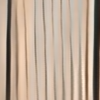}} \\ [-0.3cm]
			& & & 
			\subfloat[CARN~\cite{carn_eccv2018} \protect\linebreak(18.84/0.7132)]
			{\includegraphics[width = \patchSize, height = \patchSize]
				{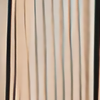}} &
			\subfloat[MSRN~\cite{msrn_eccv2018} \protect\linebreak(18.81/0.7224)]
			{\includegraphics[width = \patchSize, height = \patchSize]
				{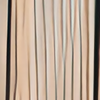}} &
			\subfloat[SHSR\protect\linebreak(\textbf{19.09}/\textbf{0.7340})]
			{\includegraphics[width = \patchSize, height = \patchSize]
				{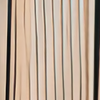}} \\
		\end{tabular}
	\end{center}
	\setlength{\abovecaptionskip}{0pt plus 2pt minus 2pt}
	\setlength{\belowcaptionskip}{0pt plus 2pt minus 2pt}
	\caption{Visual quality comparisons for various image SR methods with scaling factor $\times4$.}
	\label{fig:slogan}
\end{figure}

There are inherent correlations among the multi-scale features in the convolutioal networks. There are works focusing on visualizing and understanding the convolutional networks~\cite{review2_1, reviewer2_2}. Zeiler et al. utilized the deconvolution~\cite{review2_1} to demonstrate and investigate the information that CNN learned~\cite{reviewer2_2}. According to the visualization results, the shallow layers learn some basic features while the deeper layers focus on the discriminative features. The first and second layers in CNN learn the low-level information, such as color and edges, while the last layers concentrate on the high-level semantic features and structural information, such as heads, eyes, etc. There are also networks visualizing the learned features from shallow and deep layers~\cite{reviewer2_3, review2_4}.

In this paper, we find that larger scale features concentrate more on high-level and structural information, while the smaller scale features contain more textured information and details. In this point of view, the larger scale information can be explored from the smaller features. Based on the observation, we consider the multi-scale exploration and build the filters in a progressive way. A sequential multi-scale block (SMB) is designed to progressively explore the hierarchical information with restricted parameters. Different from exploitation with larger filters, linearity of convolution is considered and small kernels are utilized to substitute the larger ones with fewer parameters. To further investigate the inter-scale feature correlations, non-linearity is introduced between the filters. 

In addition, we build a distribution transformation block (DTB) to adjust the feature maps in a normalization manner. Different from recent attention-based methods, DTB jointly considers the channel-wise and spatial correlations, and learns both scaling and bias factor to transform the feature distribution. 
Based on the SMB and DTB, we build a sequential hierarchical learning super-resolution network (SHSR) for SISR. Experimental results show SHSR achieves better PSNR/SSIM results than state-of-the-arts. Specially, SHSR achieves superior performance with near 34\% parameters and 50\% MACs off when scaling factor is $\times4$. Besides better quantitative performance, SHSR can also restore more accurate textures than other works, as shown in Fig.~\ref{fig:slogan}. The extension model SHSR$^+$ with self-ensemble achieves competitive performance than larger networks, which holds near 92\% parameters and 42\% MACs off with scaling factor $\times4$.


Our contributions are summarized as follows:
\begin{itemize}
	\item
	We devise a sequential multi-scale block (SMB) to progressively explore the hierarchical features, which considers the inter-scale correlations and the linearity of convolution with restricted parameters.
	\item
	We devise a distribution transformation block (DTB) to adjust the feature maps in a normalization manner, which jointly considers the spatial and channel-wise correlations with scaling and bias factors.
	\item
	We build a sequential hierarchical learning super-resolution network (SHSR), which achieves superior performance to other works. The extension model SHSR$^+$ achieves competitive capacity than larger networks with much fewer parameters and lower computation complexity.
\end{itemize}

\section{Related Works}
\subsection{Deep Learning for SISR} 
SISR has proved to be a challenging issue in image restoration area, which suffers from down-sampling, blurry, and additive Gaussian white noises. Besides the complex degradation situations, different image acquisition circumstances also influence the restoration quality, such as night vision, inpainting, and moving blur. From this point of view, SISR is highly ill-posed with a large information loss.

Recently, CNN has demonstrated amazing restoration performance on SISR. SRCNN~\cite{srcnn_pami2016} proposed by Dong \textit{et al.} is the first CNN-based SISR work, which is inspired by sparse coding methods. After SRCNN, there are deeper or wider networks aiming to achieve superior capacities. FSRCNN~\cite{fsrcnn_eccv2016} avoids to upsample the input instance and builds a deeper network with fewer parameters. VDSR~\cite{vdsr_cvpr2016} investigates a very deep network to learn the residual map for SISR, which keeps the feature size by zero-padding. EDSR~\cite{edsr_cvpr2017}, MemNet~\cite{memnet_iccv2017}, RDN~\cite{rdn_pami2020}, RCAN~\cite{rcan_eccv2018}, and other latest works utilize deeper or wider networks with well-designed blocks to achieve state-of-the-art performances. There are different strategies to build effective networks. DBPN~\cite{dbpn_pami2020} designs up- and down-projection blocks following the traditional back projection method. OISR~\cite{oisr_cvpr2019} investigates an ODE-inspired block for explainable restoration. IRCNN~\cite{ircnn_cvpr2017}, DPSR~\cite{dpsr_cvpr2019}, and USRNet~\cite{usrnet_cvpr2020} regard the SISR as an optimization problem and utilize half quadratic splitting to iteratively restore the image. DRN~\cite{drn_cvpr2020} addresses the consistency between LR and down-scaled HR instances, and builds a close loop for constraining the HR and LR images based on RCAB~\cite{rcan_eccv2018}.

As one of the effective patterns, multi-scale design has been widely considered in image restoration area. There are also multi-scale networks for SISR with good capacity. LapSRN~\cite{lapsrn_cvpr2017}, MS-LapSRN~\cite{lapsrn_pami2019}, DRLN~\cite{drln_pami2020} and other Laplacian pyramid inspired networks~\cite{laplacian_tnnls2020} regard the hierarchical exploration with explicit resolution variation. Another kind of multi-scale network is to explore the feature maps with different receptive fields. MSRN~\cite{msrn_eccv2018} stacks $3\times3$ and $5\times5$ convolutional layers to exploit the multi-scale information, and utilizes residual learning for improving the gradient transmission. Recently, SMSR~\cite{smsr_tgrs2020}, MSWSR~\cite{mswsr_jstsp2020}, and other hierarchical networks also achieve state-of-the-art performance. MDCN~\cite{mdcn_csvt2020} considers more receptive field combinations and effective connection pathways. In fact, these receptive field based works seldom concentrate on the inter-scale correlations, and the larger filters usually require more parameters.

Recently, there are numerous elaborate network designs for effective image SR. Progressive restoration framework has been considered in recent works, which has been proved to be efficient~\cite{survey_pami2021}. In addition to the progressive framework, iterative super-resolution is also investigated to consider the correlations between HR and LR images~\cite{survey_pami2021}. Besides the framework, elaborative network blocks have also been investigated for better performance. To improve the vanilla residual blocks, RFANet~\cite{rfanet_cvpr2020} aggregates the learned residual features and proposes a novel RFA block to make full use of the local information. Besides the residual learning, RFANet also proposes an enhanced spatial attention mechanism for further feature exploration. MSICF~\cite{msicf_sp2021} designs a multi-scale cross module with two branches to fuse the hierarchical information. The multi-scale modules are cascaded in MSICF to build several sub-networks and process the hierarchical features.

\subsection{Attention Mechanism}
Attention mechanism is one of the effective designs to adjust the feature distribution, which learns weighting factors to concentrate more on important information. Channel-wise attention~\cite{senet_pami2020} measures the information with global pooling and devises an efficient way to explore the inter-channel correlations. Since it is a tiny but effective component, numerous SISR works considers channel-wise attention for better performance, such as RCAN~\cite{rcan_eccv2018}, IMDN~\cite{imdn_mm2019}, DRN~\cite{drn_cvpr2020}, CSFN~\cite{csfm_csvt2020} and MADNet~\cite{madnet_tcyb2020}. Besides channel-wise attention, spatial attention considers the feature correlations from another perspective. RFANet~\cite{rfanet_cvpr2020} devices an advanced spatial attention block and outperforms other works. Recently, kernel attention~\cite{kam_tomm2020} referred by Zhang~\textit{et al.} also shows good performance for image restoration. However, attention mechanism only considers adjust the distribution by using weighting factors.

\section{Methodology}

\begin{figure}[t]
	\centering
	\includegraphics[width=\linewidth]{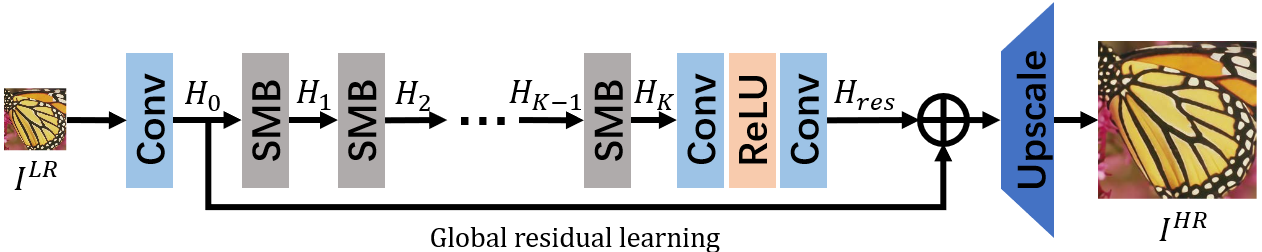}
	\caption{Illustration of the proposed SHSR.}
	\label{fig:network}
\end{figure}

\subsection{Network Structure}
Fig.~\ref{fig:network} shows the entire network design of our SHSR. Firstly, a convolutional layer extracts the features from LR image. The feature space holds more channels than the RGB space and contains more information. The feature extraction can be described as,
\begin{equation}
	H_0 = Conv(I^{LR}),
\end{equation}
where $I^{LR}$ is the LR instance, and $H_0$ is the extracted feature for hierarchical exploration.

After feature extraction, there are SMBs to explore the multi-scale information. The hierarchical features are processed in a progressive manner. For $k$-th SMB, there is,
\begin{equation}
	H_k = SMB_k(H_{k-1}),
\end{equation}
where $SMB_k(\cdot)$ denotes the $k$-th SMB separately. After $K$ SMBs, a padding structure is considered to perform the global residual learning, as,
\begin{equation}
	H_{out} = H_0 + Padding(H_K),
\end{equation}
where $Padding(\cdot)$ denotes the padding structure. The padding structure is composed of two convolutional layers and a ReLU activation. 

Finally, HR instance is restored from the explored feature by upscale block, as,
\begin{equation}
	I^{HR} = Upscale(H_{out}),
\end{equation}
where $Upscale(\cdot)$ is the upscale block and composed of two convolutional layers and a sub-pixel convolution.

\subsection{Sequential Multi-scale Block}
\begin{figure}[t]
	\captionsetup[subfloat]{labelformat=empty, justification=centering}
	\begin{center}
		\begin{tabular}[b]{c c}
			\subfloat[(a) Sequential multi-scale block.]
			{\includegraphics[width=0.5\linewidth]{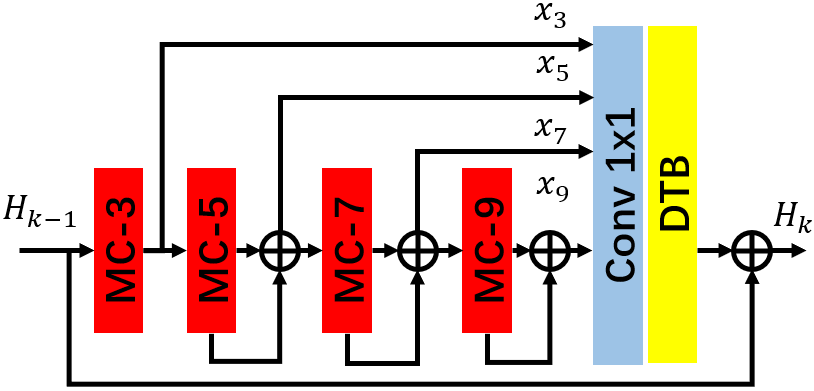}} &
			\subfloat[(b) Design of multi-scale exploration combination.]
			{\includegraphics[width=0.4\linewidth]{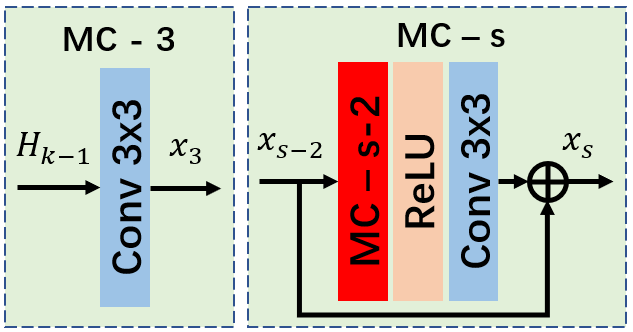}}
		\end{tabular}
	\end{center}
	\caption{Illustration of proposed sequential multi-scale block (SMB).}
	\label{fig:SMB}
\end{figure}

Fig.~\ref{fig:SMB} (a) shows the design of proposed SMB. The multi-scale combination with receptive field $s \times s$ (MC-$s$) is designed in a recursive way to explore the hierarchical information. We build the MCs in a sequential way, and derive the larger scale features from smaller ones. Let $x_s$ denotes the feature maps with scaling factor $s \times s$, then there are,
\begin{equation}
	\left\lbrace 
	\begin{aligned}
		x_{s} & = MC_s(H_{k-1}),  &s=3, \\
		x_{s} & = MC_s(x_{s-2}) + x_{s-2},  &else.
	\end{aligned}
	\right. 
	\label{eqn:SMB}
\end{equation}

From Eq.~\ref{eqn:SMB}, the larger scale features are explored from the former ones, and the sequential exploration can fully consider the inter-scale correlation of features. Instead of directly exploiting the features by filters with different receptive scales, the $MC(\cdot)$ is defined in a recursive way, as shown in Fig.~\ref{fig:SMB} (b). The first $MC_3(\cdot)$ is composed of one $3\times3$ convolutional layer. For other scaling factors, $MC_{s+2}(\cdot)$ is composed of an identical structure of $MC_{s}(\cdot)$, a ReLU activation and a $3\times3$ convolutional layer. The recursive design is motivated by the linearity of convolution, where a larger kernel can be separated into several smaller filters to keep the receptive field. Different from directly stacking the convolutions, we introduce the ReLU activation for non-linearity into MCs, aiming to build an effective exploration. The residual learning is considered for preserving the original information. It should be noticed that there is no residual connection in $MC_3(\cdot)$ since the identical addition is implied by the convolution.

After hierarchical exploration, the different features are concatenated by one $1\times1$ convolution for information aggregation and one DTB for distribution transmission, as,
\begin{equation}
	H_k = DTB(Conv([x_3, ..., x_S])) + H_{k-1},
\end{equation}
where $S$ is the largest scaling factor, and $H_k$ is the output features of $k$-th SMB. Residual learning aims to preserve the original information.

\subsection{Distribution Transformation Block}
\begin{figure}[t]
	\centering
	\includegraphics[width=0.5\linewidth]{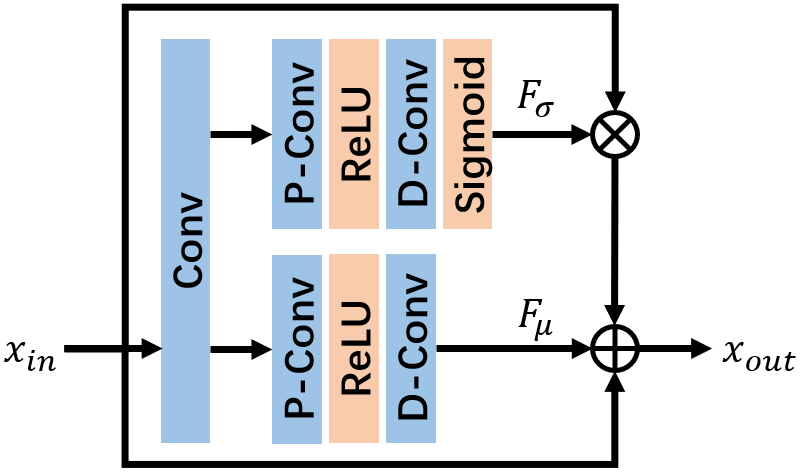}
	\caption{Illustration of the proposed distribution transformation block (DTB). Depth-wise and point-wise convolutions (D-Conv and P-Conv) are designed to explore the spatial and channel-wise correlations separately.}
	\label{fig:DTB}
\end{figure}

Fig.~\ref{fig:DTB} shows the design of proposed DTB. There is a convolutional layer for input feature $x_{in}$ to exploit the distribution information. After exploitation, depth-wise convolution (D-Conv) and point-wise convolution (P-Conv) are utilized to orthogonally explore the channel-wise and spatial correlations, and ReLU activation is utilized to introduce the non-linearity. The dual paths parallelly learn bias factor $F_\mu$ and scaling factor $F_\sigma$ separately. Sigmoid activation is specially designed for $F_\sigma$ for non-negativity.

The processing step of DTB is formulated as,
\begin{equation}
	x_{out} = (x_{in} \otimes F_\sigma + F_\mu) + x_{in},
	\label{eq:DTB}
\end{equation}
where $x_{in}$, $x_{out}$ denotes the input and output features separately, and $\otimes$ means the element-wise production. The $(x_{in} \otimes F_\sigma + F_\mu)$ in Equation~\ref{eq:DTB} means the described normalization operation of DTB. The shortcut in DTB is designed for improving the gradient transmission and preserving the original information of $x_{in}$ with local residual learning. Local residual learning has been widely considered in SISR works, and has proved to be an effective design for restoration~\cite{edsr_cvpr2017, rcan_eccv2018, rdn_pami2020, rfanet_cvpr2020}.

Eq.~\ref{eq:DTB} holds an advanced formulation than attention-based methods, which regard the bias factor $F_\mu$ as zero. Similar to normalization methods, DTB jointly utilizes bias and scaling factors to adjust the features. Traditional normalization blocks, such as batch normalization~\cite{batchnorm_icml2015} (BN), usually constrainedly adjust the feature into some standard distribution. However, this kind of adjustment may not be suitable for SISR problem which gets rid of range flexibility from networks~\cite{edsr_cvpr2017}. DTB provides the similar formulation to perform the adjustment and removes the statistical estimation step. With the learned parameters, DTB transforms the feature distribution for better SISR performance.

\subsection{Implementation Details}
In SHSR, all convolutional layers are with kernel size as $3\times3$ except for the mentioned $1\times1$ concatenation in SMB. The filter number of convolutional layers in SMBs and DTBs is set as $c=64$ while the upscale block contains $c_0=3 \times u^2$ filters, where $u$ denotes the up-sampling factor for HR image. The upscale block contains a convolutional layer with $64$ filters, and a rescaling tail that follows a similar structure as MSRN~\cite{msrn_eccv2018}, which is composed of one convolution layer and one sub-pixel convolution. There are $K=8$ SMBs stacked in SHSR, and the maximum scaling factor in SMB is set as $S=9$.

\section{Discussion}
\paragraph{Difference to MSRN~\cite{msrn_eccv2018}}
MSRN introduced a multi-scale block termed as MSRB with $3\times3$ and $5\times5$ convolutional layers. In MSRB, features from two kinds of convoluitonal layers are crossly concatenated and explored, and an $1\times1$ convolutional layer is utilized to fuse the multi-scale features.
Different from MSRB, there are features from four different scales extracted by SMB, and concatenated with one convolutional layer for fusion. Features from different scales are explored sequentially, and residual connections are utilized for information preservation.
Multi-scale information is extracted by layers with different kernel sizes in MSRB, while SMB designs the multi-scale structure in a recursive way, which focuses on the inter-scale correlations and decreases the network parameters.
Features from different MSRBs are collected and concatenated with a convolutional layer for global feature fusion, while blocks in SHSR are stacked with global residual learning.
Besides multi-scale design, a novel distribution transformation mechanism DTB is designed in SHSR. With the elaborated design, SHSR achieves better PSNR/SSIM results on all testing benchmarks than MSRN with fewer parameters and lower computation complexity.

SMB is designed based on the observation that larger scale information can derive from the smaller features. As such, the explored smaller features are helpful for effective larger scale exploration. Different from SMB, MSRN considers the hierarchical exploration in a parallel design, and utilizes the filters with different receptive fields to process the features. On one hand, the parallel design is not efficient for hierarchical exploration since it does not consider the relationship among hierarchical features. On the other hand, using filters with different receptive fields is not an efficient way to process the multi-scale information. According to the principle of convolution, the larger filter can be equivalently decomposed into a sequence of smaller filters with fewer parameters and lower computational complexity~\cite{inception_cvpr2015}. As such, our network achieves better PSNR/SSIM performance than MSRN with much fewer parameters and lower MACs. Compared with MSRN, our network achieves 0.1 dB PSNR higher on Urban100 and 0.2 dB higher on Manga109 with scaling factor $\times4$, while our network only holds near 56\% parameters and MACs.

\paragraph{Difference to Channel-wise Attention~\cite{senet_pami2020}}
There is an effective channel-wise attention design in SENet, which has been widely utilized for different image restoration problems. In channel-wise attentions, information from different channels is evaluated by global average pooling. Two full connection layers with a ReLU activation are designed to explore the attentions, and a Sigmoid activation is introduced for non-negativity.

DTB aims to learn both the channel-wise and spatial attention for image SR. There is global average pooling in vanilla channel-wise attention methods, which does not consider the spatial diversity of different feature maps. The output of channel-wise attention is a vector for weighting the feature maps, and the pixels on the same feature map share the same weight. Different from channel-wise attention, DTB considers the distribution transformation with both scaling and bias factors. The operation of DTB follows a similar formulation to the normalization manner. The D-Conv and P-Conv in DTB orthogonally explore the feature maps, and calculate the bias and scaling factors with the same shape of input features. In other words, every pixel on the feature maps will get different weights and offsets. The normalization manner and the diversity of factors make the DTB performs better than vanilla channel-wise attention. According to the ablation study, DTB achieves 0.03 dB PSNR higher than channel-wise attention on different benchmarks with scaling factor $\times4$.

\paragraph{Difference to MS-LapSRN~\cite{lapsrn_pami2019}}
MS-LapSRN is a progressive network for image super-resolution. In MS-LapSRN, the progressive structure is designed for images restorations with multiple resolutions by using one network. Residual maps are learned from the network sequentially with the increase of resolutions. In SHSR, an end-to-end network is proposed for image super-resolution with a specific up-sampling factor. The multi-scale structure is mainly designed in SMB to extract the hierarchical information. Information from multi-scale features is sequentially extracted to concentrate on the inter-scale correlations. By utilizing the sequential exploration and building a deeper network, SHSR achieves better performance than MS-LapSRN.

MS-LapSRN builds the network following the Laplacian Pyramid architecture, and progressively enlarges the feature’s resolution for multi-scale restoration. The blocks in MS-LapSRN are simply designed by stacking the convolutional and ReLU layers. To boost the restoration performance, MS-LapSRN shares the parameters of different components and restore the image in a recursive way. Different from MS-LapSRN, SMB considers the multi-scale exploration by designing the blocks with different receptive fields, rather than simply enlarging the feature map. We argue that enlarging the feature map will increase the computational cost. As such, our network achieves near 1 dB PSNR higher than MS-LapSRN with scaling factor $\times4$.

\paragraph{Difference to Recent Hierarchical Designs}
Recently, there are multi-scale designs for effective image SR. HDRN~\cite{hdrn} and HRAN~\cite{hran} are two representative hierarchical networks. The proposed sequential multi-scale block (SMB) in our network enjoys a different motivation and design from HDRN and HRAN for more effective hierarchical exploration.

HDRN stacks the convolutional layers with the same receptive field, and densely connects the features to consider the hierarchy. The dense connection in HDRN directly mixtures the explored features from all scales, which does not explicitly consider the correlation of multi-scale information. Different from HDRN, we observe that the larger scale information can derive from the smaller scale features, and design the sequential feature exploration in SMB for effective hierarchical processing. Based on the sequential design, our network achieves near 0.3 dB PSNR higher than HDRN with scaling factor $\times4$.

Similar to HDRN, HRAN considers the hierarchical exploration by concatenating the explored features from different blocks with $1\times1$ convolution. The blocks in HRAN enjoy the same network design and receptive field, and do not consider the hierarchical correlations. In our SMB, the hierarchical exploration is considered in a sequential manner, which follows the observation of the multi-scale correlation. The blocks in SMB hold different receptive fields and progressively explore the features. According to the report of HRAN~\cite{hran}, there is around 0.06 dB PSNR benefit from its hierarchical exploration. In our network, the multi-scale design in SMB provides near 0.3 dB PSNR improvement in Table~\ref{tab:abl_ms}.

\section{Experiments}

\begin{figure}[t]
	\centering
	\includegraphics[width=0.5\linewidth]{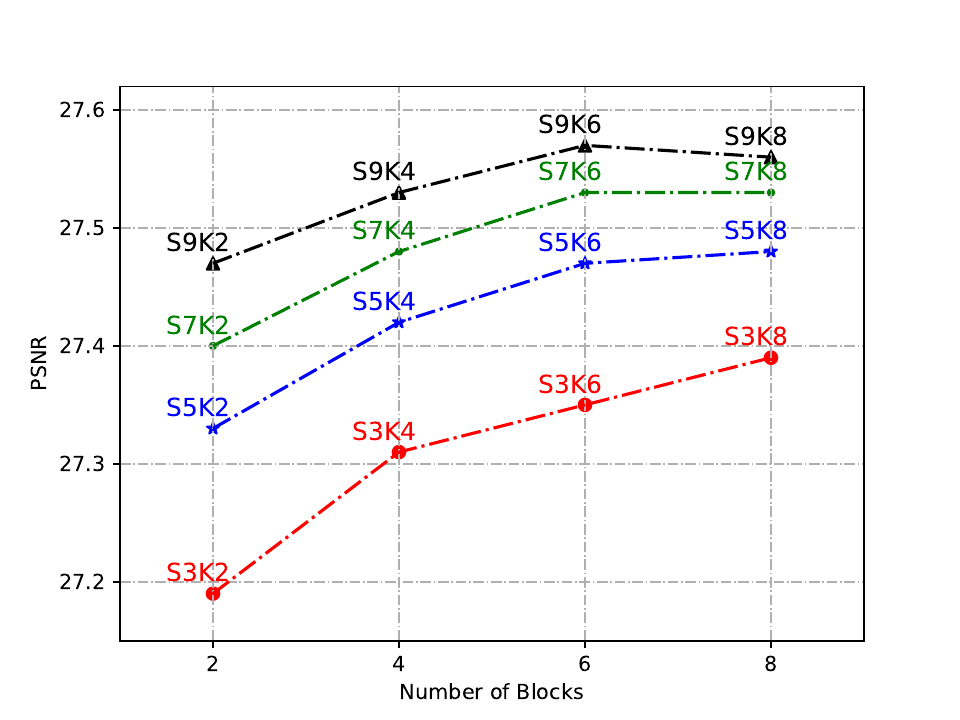}
	\caption{Investigation on different $S$ and $K$ with scaling factor $\times4$.}
	\label{fig:vis-network-depth}
\end{figure}

\subsection{Settings}
The proposed SHSR is trained with DIV2K~\cite{div2k} dataset. DIV2K is a high-quality dataset with 2K resolution images from real world. There are 800 training images, 100 validation images and 100 test images in DIV2K dataset. In this paper, 800 images are chosen for training and 5 images for validation. For testing, five benchmarks widely used in image super-resolution works: Set5~\cite{set5}, Set14~\cite{set14}, B100~\cite{b100}, Urban100~\cite{urban100}, and Manga109~\cite{manga109} are chosen. We train our SHSR with PyTorch~\cite{pytorch} on NVIDIA GTX-1080Ti GPU. The training images are randomly flipped and rotated for data augmentation. Patch size of LR image for training is set as $48\times48$. SHSR are trained for 1000 iterations with $\ell_1$ loss, and the parameters are updated with an Adam~\cite{adam} optimizer. The learning rate of optimizer is chosen as $lr=10^{-4}$, and halved for every 200 iterations. The degradation model is chosen as \textit{bicubic down}~(\textbf{BI}) with scaling factor $\times2$, $\times3$, and $\times4$. PSNR and SSIM are chosen as the indicators for quantitative comparison with other works. Self-ensemble strategy is used to improve the performance, and the extension model is termed as as SHSR$^+$.

\subsection{Model Analysis}
\subsubsection{Analysis on Network Settings}
In SHSR, the largest scale of SMB is set as $S=9$ and the number of SMB is chosen as $K=8$. To show the effect of $S$ and $K$, models are trained with different scales and block numbers for 200 epochs. Quantitative comparisons are made on B100 with scaling factor $\times4$. The visualization results are shown in Fig.~\ref{fig:vis-network-depth}. From Fig.~\ref{fig:vis-network-depth}, both $S$ and $K$ can affect the network performance. In general, with the increase of $S$ and $K$, the network achieves better results. Compared with $K$, $S$ counts more for the performance. On one hand, when $S$ is larger, the network is deeper. On the other hand, with the increase of $S$, features from more scales are considered.

\subsubsection{Analysis on SMB}
\begin{figure}[t]
	\captionsetup[subfloat]{labelformat=empty, justification=centering}
	\begin{center}
		\scriptsize
		\setlength\tabcolsep{0.1cm}
		\begin{tabular}[b]{cccccc}
			\subfloat[(a)]{\includegraphics[width=0.15\linewidth]{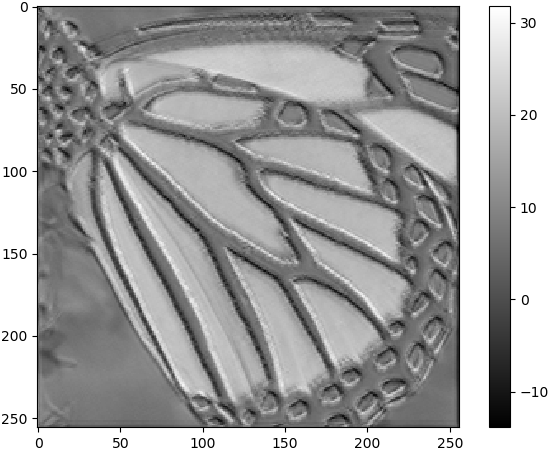}}&
			\subfloat[(b)]{\includegraphics[width=0.15\linewidth]{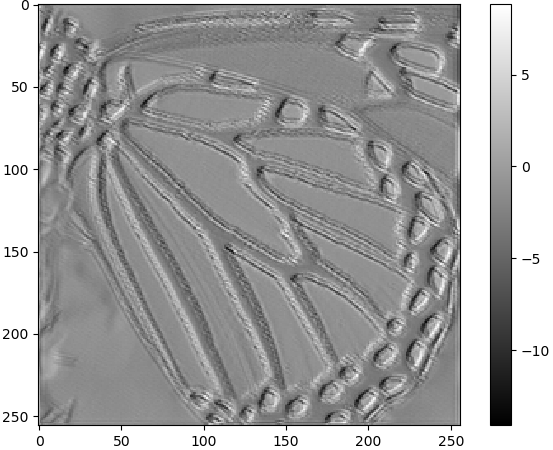}}&
			\subfloat[(c)]{\includegraphics[width=0.15\linewidth]{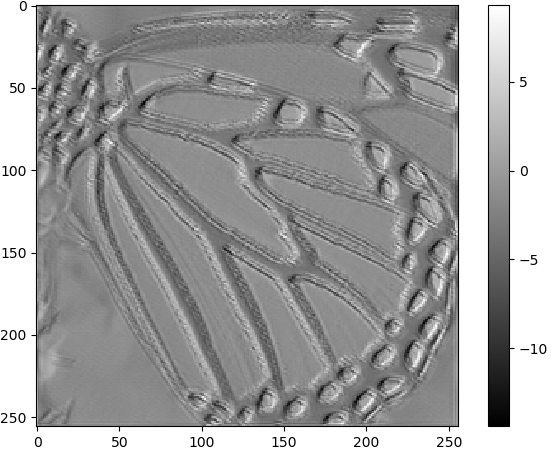}}&
			\subfloat[(d)]{\includegraphics[width=0.15\linewidth]{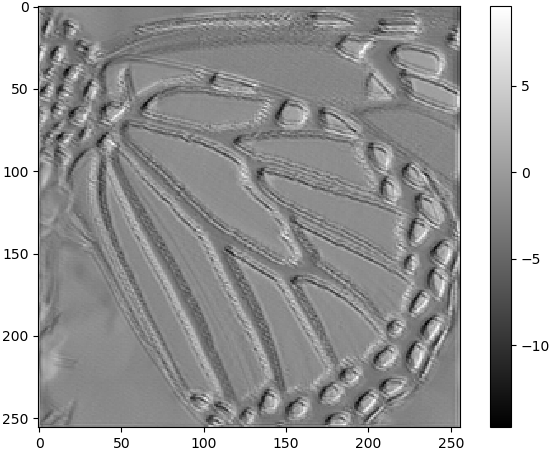}}&
			\subfloat[(e)]{\includegraphics[width=0.15\linewidth]{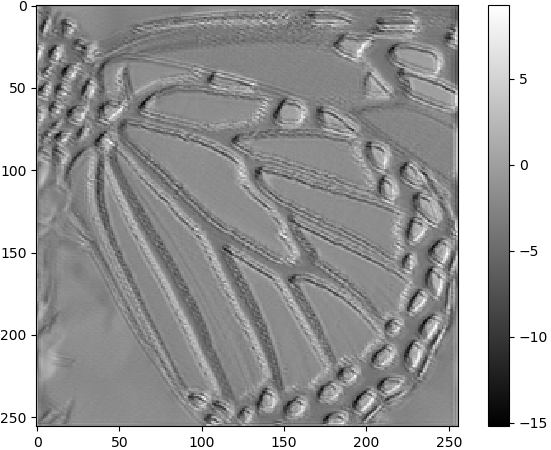}}&
			\subfloat[(f)]{\includegraphics[width=0.15\linewidth]{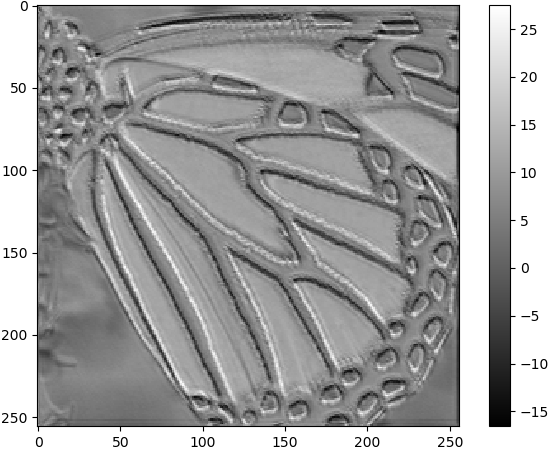}}
		\end{tabular}
	\end{center}
	\setlength{\abovecaptionskip}{0pt plus 2pt minus 2pt}
	\setlength{\belowcaptionskip}{0pt plus 2pt minus 2pt}
	\caption{Illustrations of multi-scale features. (a) and (f) denote the input and output features. (b)-(e) denote the features with scaling factor 3, 5, 7, and 9.}
	\label{fig:feat}
\end{figure}
\begin{table}[t]
	\centering
	\caption{PSNR/SSIM results of multi-scale mechanism in SMB with scaling factor $\times4$.}
	\label{tab:abl_ms}
	\begin{tabular}{|c|c|c|c|c|}
		\hline  
		\textbf{Multi}& \textbf{Set5}& \textbf{Set14}& \textbf{B100} &\textbf{Urban100}\\
		\hline
		\hline
		{w}& 32.34/0.8971& 28.71/0.7850& 27.66/0.7392& 26.37/0.7953\\ 
		{w/o}& 32.03/0.8932& 28.51/0.7799& 27.53/0.7348& 25.90/0.7803\\
		\hline
	\end{tabular}
\end{table}
There is multi-scale design in SMB extracting the hierarchical information. To show the performance of multi-scale design, comparisons are conducted without different combinations of convolutional layers. All combinations are replaced by only one $3\times3$ convolutional layer. In other words, all the scales in SMB are identical to $3\times3$. The results are shown in Table.~\ref{tab:abl_ms} on four benchmarks with scaling factor $\times4$.
From Table~\ref{tab:abl_ms}, model with multi-scale design achieves better PSNR/SSIM results than the other one. There are two reasons for the performance improvement. On one hand, the features of different scales contain more information, which helps to recover the complex structural textures. On the other hand, the multi-scale structures are built in a recursive way. With the combination of convolutional layers, the depth of SHSR is increased, which may be helpful to improve the network representation.

Furthermore, we analyze the exploited features from different scales, which are shown in Fig.~\ref{fig:feat}. The multi-scale features are exploited from different layer combinations. With the increasing of scale factors, the structural information becomes sharper and clearer, and the tiny textures become flat. This accords with the notion that multi-scale features contain different information.

In SMB, residual connections are introduced to preserve the information from small scales. Feature fusion with $1\times1$ convolution is also used to concatenate information from different scales. To show the performance of information preservation and feature fusion, we perform the comparisons without residual and $1\times1$ convolution. The results are shown in Table.~\ref{tab:abl_pmrb}, where \textbf{Res} and \textbf{Fuse} denote the residual connection and concatenation separately. Three benchmarks covering different kinds of textures are used for testing with scaling factor $\times4$.
From the Table~\ref{tab:abl_pmrb}, residual and feature fusion are both efficient for different benchmarks. For Set5, residual structure performs better than fusion, achieving around 0.1dB improvement. For B100 and Urban100, feature fusion can recover the texture more effectively. Set5 contains less high-frequency information than the other benchmarks, while B100 and Urban100 are composed of abundant images from real world. From this perspective, residual connection is suitable for simple images, while feature fusion performs better on complex structural textures.

\begin{table}[t]
	\centering
	\caption{PSNR/SSIM results of different structures in SMB with scaling factor $\times4$.}
	\label{tab:abl_pmrb}
	\begin{tabular}{|c|c|c|c|c|}
		\hline  
		\textbf{Res}& \textbf{Fuse}& \textbf{Set5}& \textbf{B100} &\textbf{Urban100} \\
		\hline
		\hline
		w&w& 32.34/0.8971& 27.66/0.7392& 26.37/0.7953\\ 
		w&w/o& 32.35/0.8971& 27.64/0.7384& 26.34/0.7942\\
		w/o&w& 32.24/0.8963& 27.65/0.7388& 26.36/0.7955\\
		\hline
	\end{tabular}
\end{table}

\subsubsection{Analysis on Multi-scale Combination.}
In SHSR, recursive layer combinations (MCs) are proposed to substitute convolutional layers with different kernel sizes. To show the performance of substitution, PSNR/SSIM comparisons are made on five benchmarks with scaling factor $\times4$. To ensure the same receptive field, network without combinations is built with layers holding the kernel sizes as $5\times5$, $7\times7$ and $9\times9$ separately. The results are shown in Table~\ref{tab:abl_3x3}.
From Table~\ref{tab:abl_3x3}, model built with layer combinations achieves better PSNR/SSIM results on all five testing benchmarks, showing the performance of recursive design. Meanwhile, there are around 40.2\% off on parameters and MACs when utilizing recursive combinations. 

\begin{table*}
	\centering
	\caption{PSNR/SSIM results, parameters, and MACs of recursive combination in SMB with scaling factor $\times4$.}
	\label{tab:abl_3x3}
	\begin{tabular}{|c|c|c|c|c|c|c|c|}
		\hline  
		\textbf{Comb}&\textbf{Param}& \textbf{MACs}& \textbf{Set5}& \textbf{Set14}& \textbf{B100} &\textbf{Urban100}& \textbf{Manga109}\\
		\hline
		\hline
		{w}& 3,598K& 207.2G&32.34/0.8971& 28.71/0.7850& 27.66/0.7392& 26.37/0.7953& 30.71/0.9107\\ 
		{w/o}& 6,020K& 346.7G&32.07/0.8932& 28.53/0.7804& 27.53/0.7350& 25.93/0.7819& 30.16/0.9043\\
		\hline
	\end{tabular}
\end{table*}

\subsubsection{Analysis on Distribution Transformation.}
In SHSR, DTB is investigated for distribution transformation. To show the performance of proposed DTB, comparisons are designed on three testing benchmarks. We compare the models with DTB, channel-wise attention~(CA)~\cite{senet_pami2020}, and no transformation. The results are shown in Table~\ref{tab:abl_norm}.
From the table, the model with DTB achieves the best performance on all testing benchmarks. The model with channel-wise attentions achieves better PSNR/SSIM results than that without attentions. The results demonstrate that distribution transformation is efficient for image super-resolution.

\begin{table}[t]
	\centering
	\caption{PSNR/SSIM results of different transformations with scaling factor $\times4$.}
	\label{tab:abl_norm}
	\begin{tabular}{|c|c|c|c|}
		\hline  
		\textbf{Method}& \textbf{Set5}& \textbf{Set14}& \textbf{Urban100}\\
		\hline
		\hline
		{DTB}	& 32.34/0.8971& 28.71/0.7850& 26.37/0.7953\\ 
		{CA~\cite{senet_pami2020}}	& 32.31/0.8968& 28.69/0.7844& 26.34/0.7940\\
		{w/o}		& 32.29/0.8965& 28.68/0.7851& 26.29/0.7940\\
		\hline
	\end{tabular}
\end{table}

\begin{figure}[t]
	\captionsetup[subfloat]{labelformat=empty, justification=centering}
	\begin{center}
		\begin{tabular}[b]{cccc}
			\subfloat[(a)~Feature before DTB]{\includegraphics[width=0.23\linewidth]{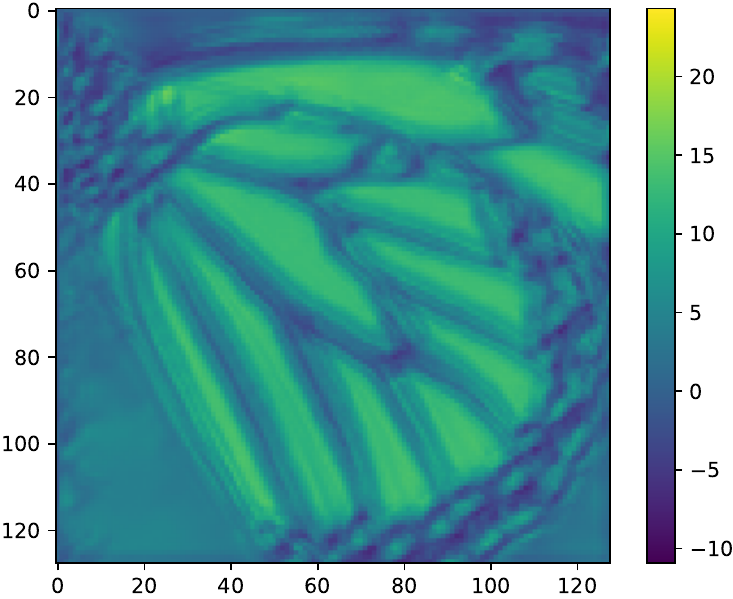}}&
			\subfloat[(b)~Feature after DTB]{\includegraphics[width=0.23\linewidth]{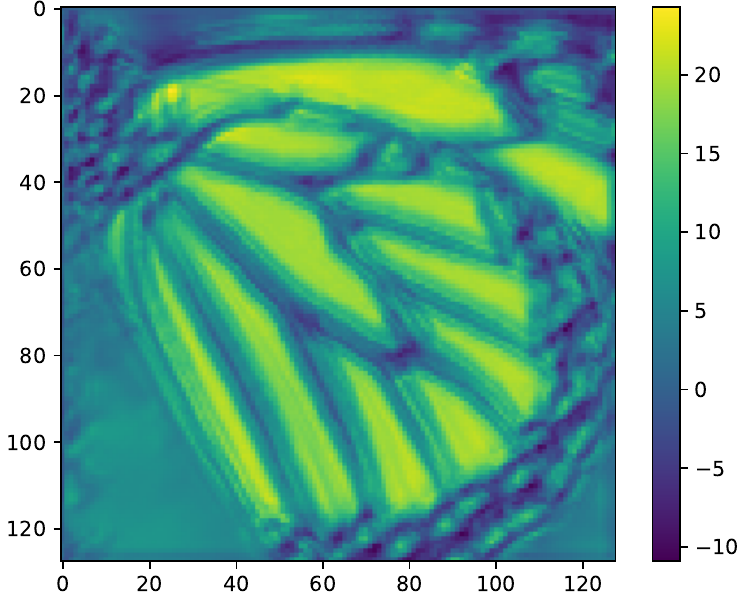}}&
			\subfloat[(c)~$F_\sigma$ from DTB]{\includegraphics[width=0.23\linewidth]{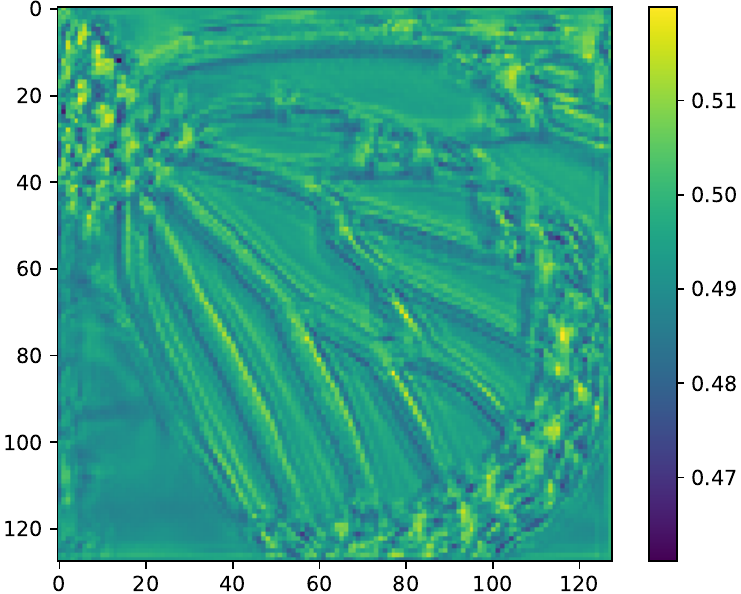}}&
			\subfloat[(d)~$F_\mu$ from DTB]{\includegraphics[width=0.23\linewidth]{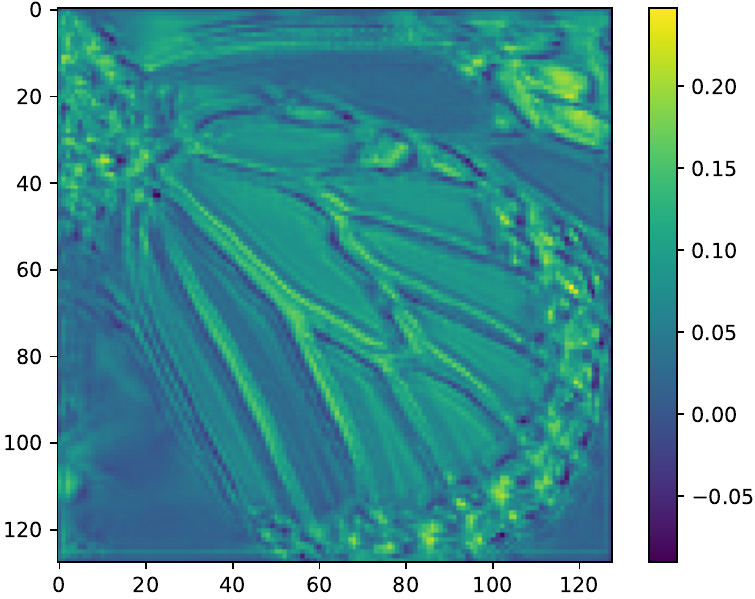}}\\
		\end{tabular}
	\end{center}
	\caption{Visualization feature maps of DTB.}
	\label{fig:vis-attention}
\end{figure}

To analyze the operation of DTB, factors $F_\gamma$, $F_\beta$ and the feature maps before and after attention are visualized in Fig.~\ref{fig:vis-attention}. From the illustrations, learned attentions are more concentrated on structural textures. $F_\gamma$ and $F_\beta$ vary sharply on the area of edges and complex textures. After attentions, the features are more discriminative on structural textures, which is a convincing evidence of that the transformation mechanism concentrates more on the important high-frequency information.

\subsection{Comparison with State-of-the-Arts}
\begin{table*}[t]
	\centering
	\caption{Average PSNR/SSIM, parameters and MACs results with degradation model \textbf{BI} $\times2$, $\times3$, and $\times4$ on five benchmarks. The best and second performances are shown in \textbf{bold} and \underline{underline}.}
	\label{tab:BI-result}
	\fontsize{7.5}{8}\selectfont
	\begin{tabular}{|c|c|c|c|c|c|c|c|c|}
		\hline
		\multirow{2}{*}{Scale}& \multirow{2}{*}{Model}&  \multirow{2}{*}{Params}& \multirow{2}{*}{MACs}& Set5~\cite{set5}& Set14~\cite{set14}& B100~\cite{b100}& Urban100~\cite{urban100}& Manga109~\cite{manga109} \\
		& & & & PSNR/SSIM & PSNR/SSIM & PSNR/SSIM & PSNR/SSIM & PSNR/SSIM\\
		\hline
		\hline
		\multirow{17}{*}{$\times2$} &SRCNN~\cite{srcnn_pami2016}&57K &52.7G & 
		36.66/0.9542& 32.42/0.9063& 31.36/0.8879& 29.50/0.8946& 35.74/0.9661\\
		
		& FSRCNN~\cite{fsrcnn_eccv2016}&12K&6.0G& 
		37.00/0.9558& 32.63/0.9088& 31.53/0.8920& 29.88/0.9020& 36.67/0.9694\\
		
		& VDSR~\cite{vdsr_cvpr2016}&665K&612.6G& 
		37.53/0.9587& 33.03/0.9124& 31.90/0.8960& 30.76/0.9140& 37.22/0.9729\\
		
		& DRCN~\cite{drcn_cvpr2016}&1,774K&17,974.3G& 
		37.63/0.9588& 33.04/0.9118& 31.85/0.8942& 30.75/0.9133& 37.63/0.9723\\
		
		& CNF~\cite{cnf_cvprw2017}&337K&311.0G& 
		37.66/0.9590& 33.38/0.9136& 31.91/0.8962& - & - \\
		
		&LapSRN~\cite{lapsrn_cvpr2017}&813K&29.9G& 
		37.52/0.9590& 33.08/0.9130& 31.80/0.8950& 30.41/0.9100& 37.27/0.9740\\
		
		&DRRN~\cite{drrn_cvpr2017}&297K&6,796.9G& 
		37.74/0.9591& 33.23/0.9136& 32.05/0.8973& 31.23/0.9188& 37.92/0.9760\\
		
		&BTSRN~\cite{btsrn_cvprw2017}&410K&207.7G& 
		37.75/-& 33.20/-& 32.05/-& 31.63/-& -\\
		
		&MemNet~\cite{memnet_iccv2017}&677K&2,662.4G& 
		37.78/0.9597& 33.28/0.9142& 32.08/0.8978& 31.31/0.9195& 37.72/0.9740 \\
		
		&SelNet~\cite{selnet_cvprw2017}&974K&225.7G& 
		37.89/0.9598& 33.61/0.9160& 32.08/0.8984& - &  - \\
		
		&CARN~\cite{carn_eccv2018}&1,592K&222.8G& 
		37.76/0.9590& 33.52/0.9166& 32.09/0.8978& 31.92/0.9256& 38.36/0.9765\\
		
		&MSRN~\cite{msrn_eccv2018}& 5,930K& 1367.5G& 
		38.08/0.9607& 33.70/0.9186& 32.23/0.9002& 32.29/0.9303& \underline{38.69/0.9772} \\
		
		&PRLSR~\cite{prlsr_icassp2020}& 1,363K& 187.3G&
		38.09/0.9608& 33.69/.09191& 32.25/0.9005& 32.35/0.9308& - \\
		
		&OISR-RK2~\cite{oisr_cvpr2019}& 4,970K& 1145.7G&
		\underline{38.12/0.9609}& \underline{33.80/0.9193}& \underline{32.26/0.9006}& \underline{32.48/0.9317}& - \\
		
		&LMAN-s~\cite{lman_tob2020}& 1,531K& 91.7G&
		37.94/0.9603& 33.49/0.9167& 32.08/0.8984& 31.85/0.9251& 38.43/0.9765 \\
		
		&SHSR& 3,577K& 824.2G& 
		\textbf{38.13/0.9609}& \textbf{33.85/0.9204}& \textbf{32.28/0.9010}& \textbf{32.59/0.9328}& \textbf{38.91/0.9775} \\
		
		\cline{2-9}
		
		&EDSR~\cite{edsr_cvpr2017}&40,729K& 9,388.8G& 
		38.11/0.9602& \textbf{33.92}/0.9195& \underline{32.32/0.9013}& \textbf{32.93/0.9351}& \underline{39.10/0.9773}\\
		
		&D-DBPN~\cite{dbpn_pami2020}& 5,953K& 3,746.2G& 
		38.09/0.9600& 33.85/0.9190& 32.27/0.9000& 32.55/0.9324& 38.89/0.9775\\
		
		&SRFBN~\cite{srfbn_cvpr2019}& 2,140K& 5,043.5G& 
		\underline{38.11/0.9609}& 33.82/\underline{0.9196}& 32.29/0.9010& 32.62/0.9328& 39.08/0.9779\\
		
		
		&SHSR$^+$& 3,577K& 6,593.6G&
		\textbf{38.22/0.9612}& \underline{33.90}/\textbf{0.9205}& \textbf{32.34/0.9015}& \underline{32.78/0.9342}& \textbf{39.15/0.9781} \\
		\hline
		\hline
		\multirow{16}{*}{$\times3$}& SRCNN~\cite{srcnn_pami2016} &57K &52.7G & 
		32.75/0.9090& 29.28/0.8209& 28.41/0.7863& 26.24/0.7989& 30.59/0.9107\\
		
		&FSRCNN~\cite{fsrcnn_eccv2016}&12K&5.0G& 
		33.16/0.9140& 29.43/0.8242& 28.53/0.7910& 26.43/0.8080& 30.98/0.9212\\
		
		&VDSR~\cite{vdsr_cvpr2016}&665K&612.6G& 
		33.66/0.9213& 29.77/0.8314& 28.82/0.7976& 27.14/0.8279& 32.01/0.9310\\
		
		&DRCN~\cite{drcn_cvpr2016}&1,774K&17,974.3G& 
		33.82/0.9226& 29.76/0.8311& 28.80/0.7963& 27.15/0.8276& 32.31/0.9328\\
		
		&CNF~\cite{cnf_cvprw2017}&337K&311.0G& 
		33.74/0.9226& 29.90/0.8322& 28.82/0.7980& - & - \\
		
		&DRRN~\cite{drrn_cvpr2017}&297K&6,796.9G& 
		34.03/0.9244& 29.96/0.8349& 28.95/0.8004& 27.53/0.8378& 32.74/0.9390\\
		
		&BTSRN~\cite{btsrn_cvprw2017}&410K&176.2G& 
		34.03/-& 29.90/-& 28.97/-& 27.75/-& -\\
		
		&MemNet~\cite{memnet_iccv2017}&677K&2,662.4G& 
		34.09/0.9248& 30.00/0.8350& 28.96/0.8001& 27.56/0.8376& 32.51/0.9369 \\
		
		&SelNet~\cite{selnet_cvprw2017}&1,159K&120.0G& 
		34.27/0.9257& 30.30/0.8399& 28.97/0.8025& - &  - \\
		
		&CARN~\cite{carn_eccv2018}&1,592K&118.8G& 
		34.29/0.9255& 30.29/0.8407& 29.06/0.8034& 28.06/0.8493& 33.49/0.9440\\
		
		&MSRN~\cite{msrn_eccv2018}& 6,114K& 626.6G& 
		34.46/0.9278& 30.41/0.8437& 29.15/0.8064& 28.33/0.8561& \underline{33.67/0.9456} \\
		
		&OISR-RK2~\cite{oisr_cvpr2019}& 5,640K& 578.6G&
		\underline{34.55/0.9282}& \textbf{30.46}/0.8443& \underline{29.18/0.8075}& \underline{28.50/0.8597}& -\\
		
		&PRLSR~\cite{prlsr_icassp2020}& 1,456K& 94.5G&
		34.47/0.9278& 30.43/0.8436& 29.14/0.8060& 28.27/0.8541& - \\
		
		&SHSR&3,586K& 366.6G& 
		\textbf{34.57/0.9284}& \underline{30.43}/\textbf{0.8444}& \textbf{29.19/0.8075}& \textbf{28.51/0.8601}& \textbf{33.85/0.9465}  \\
		
		\cline{2-9}
		
		&EDSR~\cite{edsr_cvpr2017}&43,680K& 4,471.5G& 34.65/0.9280& \underline{30.52}/\textbf{0.8462}& \textbf{29.25/0.8093}& \textbf{28.80/0.8653}& \underline{34.17/0.9476}\\
		
		&SRFBN~\cite{srfbn_cvpr2019}&2,832K& 6,023.8G& \textbf{34.70/0.9292}& 30.51/0.8461& 29.24/0.8084& \underline{28.73/0.8641}& \textbf{34.18/0.9481}\\
		
		
		&SHSR$^+$&3,586K& 2,932.8G&
		\underline{34.65/0.9289}& \textbf{30.54}/\underline{0.8461}& \underline{29.24/0.8087}& 28.71/0.8630& 34.10/0.9480 \\
		\hline
		\hline
		\multirow{18}{*}{$\times4$}&SRCNN~\cite{srcnn_pami2016}&57K &52.7G & 
		30.48/0.8628& 27.49/0.7503& 26.90/0.7101& 24.52/0.7221& 27.66/0.8505\\
		
		&FSRCNN~\cite{fsrcnn_eccv2016}&12K&4.6G& 
		30.71/0.8657& 27.59/0.7535& 26.98/0.7150& 24.62/0.7280& 27.90/0.8517\\
		
		&VDSR~\cite{vdsr_cvpr2016}&665K&612.6G& 
		31.35/0.8838& 28.01/0.7674& 27.29/0.7251& 25.18/0.7524& 28.83/0.8809\\
		
		&DRCN~\cite{drcn_cvpr2016}&1,774K&17,974.3G& 
		31.53/0.8854& 28.02/0.7670& 27.23/0.7233& 25.14/0.7510& 28.98/0.8816\\
		
		&CNF~\cite{cnf_cvprw2017}&337K&311.0G& 
		31.55/0.8856& 28.15/0.7680& 27.32/0.7253& - & - \\
		
		&LapSRN~\cite{lapsrn_cvpr2017}&813K&149.4G& 
		31.54/0.8850& 28.19/0.7720& 27.32/0.7280& 25.21/0.7560& 29.09/0.8845\\
		
		&DRRN~\cite{drrn_cvpr2017}&297K&6,796.9G& 
		31.68/0.8888& 28.21/0.7720& 27.38/0.7284& 25.44/0.7638& 29.46/0.8960\\
		
		&BTSRN~\cite{btsrn_cvprw2017}&410K&207.7G& 
		31.85/-& 28.20/-& 27.47/-& 25.74/-& -\\
		
		&MemNet~\cite{memnet_iccv2017}&677K&2,662.4G& 
		31.74/0.8893& 28.26/0.7723& 27.40/0.7281& 25.50/0.7630& 29.42/0.8942 \\
		
		&SelNet~\cite{selnet_cvprw2017}&1,417K&83.1G& 
		32.00/0.8931& 28.49/0.7783& 27.44/0.7325& - &  - \\
		
		&CARN~\cite{carn_eccv2018}&1,592K&90.9G& 
		32.13/0.8937& 28.60/0.7806& 27.58/0.7349& 26.07/0.7837& 30.40/0.9082\\
		
		&MSRN~\cite{msrn_eccv2018}&6,373K& 368.6G& 
		32.26/0.8960& 28.63/0.7836& 27.61/0.7380& 26.22/0.7911& \underline{30.57/0.9103} \\
		
		& s-LWSR$_{64}$~\cite{s_LWSR_tip2020}& 2,277K& 131.1G&
		32.28/0.8960& 28.34/0.7800& 27.61/0.7380& 26.19/0.8910& - \\
		
		&OISR-RK2~\cite{oisr_cvpr2019}&5,500K&412.2G&
		\underline{32.32/0.8965}& \textbf{28.72}/\underline{0.7843}& \underline{27.66/0.7390}& \underline{26.37/0.7953}& - \\
		
		&PRLSR~\cite{prlsr_icassp2020}& 1,437K& 66.9G&
		32.31/0.8962& 28.71/0.7838& 27.64/0.7378& 26.22/0.7892& - \\
		
		&SHSR&3,598K& 207.2G& 
		\textbf{32.34/0.8971}& \underline{28.71}/\textbf{0.7850}& \textbf{27.66/0.7392}& \textbf{26.37/0.7953}& \textbf{30.71/0.9107}  \\
		
		\cline{2-9}
		
		&EDSR~\cite{edsr_cvpr2017}&43,089K& 2,895.8G& 32.46/0.8968& 28.80/\textbf{0.7876}& 27.71/\textbf{0.7420}& \textbf{26.64/0.8033}& 31.02/\underline{0.9148}\\
		
		&D-DBPN~\cite{dbpn_pami2020}&10,426K& 5,213.0G& 32.47/0.8980& \textbf{28.82}/0.7860& 27.72/0.7400& 26.38/0.7946&30.91/0.9137\\
		
		&SRFBN~\cite{srfbn_cvpr2019}&3,631K& 7,466.1G& \underline{32.47/0.8983}& 28.81/0.7868& \textbf{27.72}/\underline{0.7409}& \underline{26.60/0.8015}& \textbf{31.15/0.9160}\\
		
		
		&SHSR$^+$&3,598K& 1,657.6G&
		\textbf{32.47/0.8984}& \underline{28.81/0.7870}& \underline{27.72}/0.7405& 26.55/0.7995& \underline{31.07}/0.9144 \\
		\hline
		
	\end{tabular}
\end{table*}

To make quantitive comparison, we compare the PSNR/SSIM results with several small works: SRCNN~\cite{srcnn_pami2016}, FSRCNN~\cite{fsrcnn_eccv2016}, VDSR~\cite{vdsr_cvpr2016}, DRCN~\cite{drcn_cvpr2016}, CNF~\cite{cnf_cvprw2017}, LapSRN~\cite{lapsrn_cvpr2017}, DRRN~\cite{drrn_cvpr2017}, BTSRN~\cite{btsrn_cvprw2017}, MemNet~\cite{memnet_iccv2017}, SelNet~\cite{selnet_cvprw2017}, CARN~\cite{carn_eccv2018}, MSRN~\cite{msrn_eccv2018}, s-LWSR~\cite{s_LWSR_tip2020}, PRLSR~\cite{prlsr_icassp2020}, LMAN-s~\cite{lman_tob2020} and OISR~\cite{oisr_cvpr2019}. 
For a fair comparison, extension model SHSR$^+$ with self-ensemble is compared with large networks: EDSR~\cite{edsr_cvpr2017}, D-DBPN~\cite{dbpn_pami2020}, and SRFBN~\cite{srfbn_cvpr2019}. 

Table~\ref{tab:BI-result} shows the PSNR/SSIM comparisons among several methods. From the results, SHSR achieves competitive or better performance than other small works on all five benchmarks. Compared with the multi-scale method MSRN, SHSR gains 0.3dB increase on Urban100 with $\textbf{BI}\times2$ degradation. Furthermore, SHSR achieves competitive or better performance than OISR with around 34\% parameters off. Notice that SHSR achieves the best performances on B100, Urban100, and Manga109 with all degradation models. The three benchmarks contain plentiful structural information and edges, which consist of comic covers and real world photos. From this point of view, SHSR can recover the high-frequency information more effectively than others.

Meanwhile, we compare the computation complexity and parameters with other works. 
Computation complexity is modeled as the number of multiply-accumulate operations~(MACs). Since it is an implementation independent factor, MACs can purely describe the computation complexity from the mathematical perspective. Comparisons of MACs are conducted by producing a 720P~($1280\times720$) resolution image from corresponding LR image with different scaling factors. The MACs of SHSR is calculated by PyTorch-OpCounter~\footnote{https://github.com/Lyken17/pytorch-OpCounter} and multiplied by 8 for SHSR$^+$.

From the results, SHSR achieves competitive or better PSNR/SSIM results than others with fewer parameters and MACs, which proves to be the efficient design. Compared with OISR, SHSR holds near half MACs and parameters with competitive or better PSNR/SSIM performances with $\textbf{BI}\times4$ degradation. Compared with larger networks, SHSR$^+$ achieves competitive performances with much fewer MACs and parameters. Specially, SHSR$^+$ holds near half of the MACs and one tenth of the parameters than EDSR with $\textbf{BI}\times4$ degradation, and achieves superior PSNR results on Set5, Set14, B100, and Manga109 datasets.

To further investigate the effectiveness of SHSR, we compare our network with recent state-of-the-art deep SISR networks. Table~\ref{tab:sota} shows the PSNR, parameters and MACs comparison with scaling factor $\times4$. In the table, we mainly compare our network with recent state-of-the-art SISR works (RCAN~\cite{rcan_eccv2018}, MSRN~\cite{msrn_eccv2018}, USRNet~\cite{usrnet_cvpr2020}, DRN~\cite{drn_cvpr2020}, SRFBN~\cite{srfbn_cvpr2019}, EBRN~\cite{ebrn_iccv2019}, OISR-RK2~\cite{oisr_cvpr2019}). Our network holds near 20\% parameters and 22\% MACs than RCAN with only less than 0.5 dB PSNR degradation. Specifically, we only drop less than 0.2 dB PNSR on Set14 and B100, which is marginal for visual quality. When compared with other state-of-the-art works, our network also achieves competitive performance with much fewer parameters and MACs. There is only less than 0.2 dB PSNR drop for our network when compared with SRFBN~\cite{srfbn_cvpr2019}, while we save near 97.2\% MACs. For DRN~\cite{drn_cvpr2020}, we save near 73\% parameters and 85\% MACs with less than 0.4 dB PSNR drop. When compared with EBRN~\cite{ebrn_iccv2019}, our network saves near 70\% parameters and 97\% MACs and drops less than 0.4 dB on PSNR. MSRN~\cite{msrn_eccv2018} and OISR-RK2~\cite{oisr_cvpr2019} are two recent works with similar parameters and MACs to our network. Compared with OISR, our network achieves competitive performance and saves near 34\% parameters and 50\% MACs. Our network also achieves better PSNR results than MSRN with fewer parameters and MACs.

\begin{table*}[t]
	\centering
	\caption{PSNR, parameters and MACs comparisons with state-of-the-art SISR works with scaling factor $\times4$.}
	\label{tab:sota}
	\fontsize{6.5}{8}\selectfont
	\begin{tabular}{|c|c|c|c|c|c|c|c||c|}
		\hline
		\textbf{Method}& \textbf{RCAN~\cite{rcan_eccv2018}}& \textbf{USRNet~\cite{usrnet_cvpr2020}}& \textbf{DRN~\cite{drn_cvpr2020}}& \textbf{MSRN~\cite{msrn_eccv2018}} & \textbf{SRFBN~\cite{srfbn_cvpr2019}}& \textbf{EBRN~\cite{ebrn_iccv2019}}& \textbf{OISR-RK2~\cite{oisr_cvpr2019}}& \textbf{Ours} \\
		\hline
		\hline
		\textbf{Params (K)}& 17410& 17016& 9825& 6373& 3630& 11279& 	  5500		&3598 \\
		\textbf{MACs (G)}& 919.1& 8545.8& 1406.3& 368.6& 7466.1& 6439.0&  412.2		&207.2 \\
		\hline
		\hline
		\textbf{Set5}& 32.63& 32.42& 32.74& 32.26& 32.47& 32.79& 		  32.32		&32.34 \\
		\textbf{Set14}& 28.87& 28.83& 28.98& 28.63& 28.81& 29.01& 		  28.72		&28.71 \\
		\textbf{B100}& 27.77& 27.69& 27.83& 27.61& 27.72& 27.85& 		  27.66		&27.66 \\
		\hline
	\end{tabular}
\end{table*}

\begin{figure}[t]
	\centering
	\includegraphics[width=0.5\linewidth]{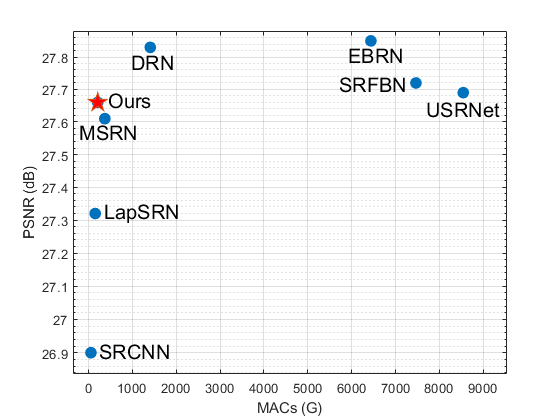}
	\caption{Visualization comparisons on PSNR and MACs with scaling factor $\times4$.}
	\label{fig:psnr-mac}
\end{figure}

\begin{figure}[t]
	\centering
	\includegraphics[width=0.5\linewidth]{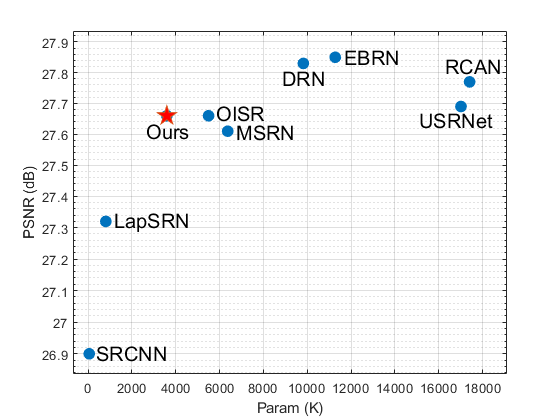}
	\caption{Visualization comparisons on PSNR and parameters with scaling factor $\times4$.}
	\label{fig:psnr-param}
\end{figure}

For better illustrating the performance, we plot the PSNR, MACs and parameter comparisons on B100 benchmark with \textbf{BI}$\times4$ degradation. Figure~\ref{fig:psnr-mac}~and~\ref{fig:psnr-param} show the comparison. From the results, our network makes a good trade-off on model complexity and performance.

\begin{figure*}[t]
	\captionsetup[subfloat]{labelformat=empty, justification=centering}
	\begin{center}
		\newcommand{\rowArg}{2.2cm}
		\newcommand{\fullSizeHei}{4.9cm}
		\newcommand{\fullSizeWid}{6cm}
		\newcommand{\patchSize}{2cm}
		\scriptsize
		\setlength\tabcolsep{0.05cm}
		\vspace{-0.3cm}
		\begin{tabular}[b]{ccccc}
			\multirow{2}{*}[\rowArg]{
				\subfloat[image\_059 from Urban100]
				{\includegraphics[width = \fullSizeWid, height = \fullSizeHei]
					{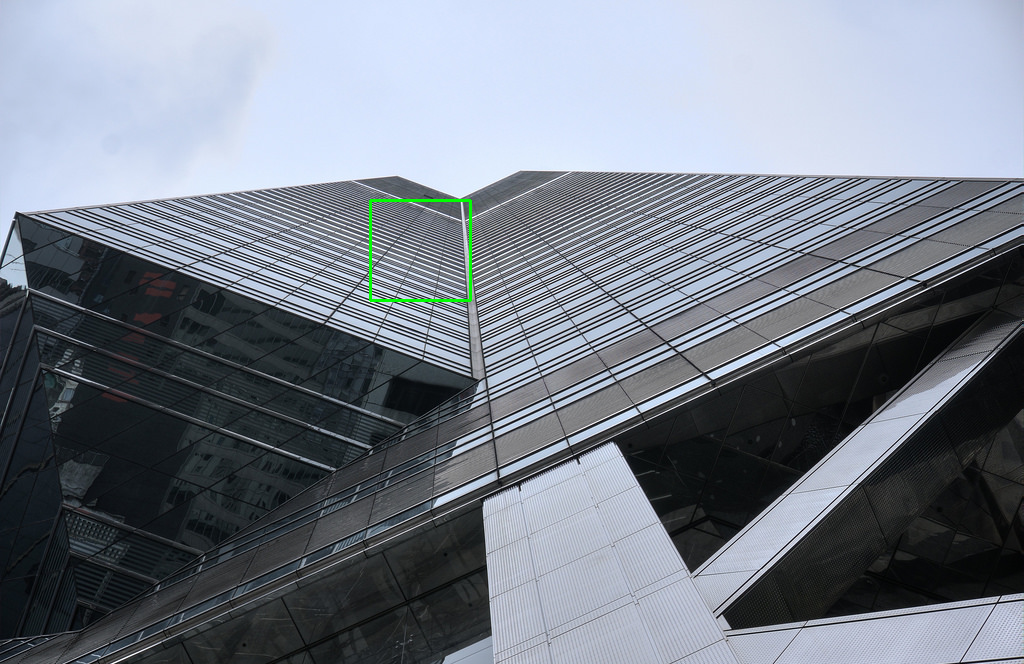}}} &
			\subfloat[HR~\protect\linebreak(PSNR/SSIM)]
			{\includegraphics[width = \patchSize, height = \patchSize]
				{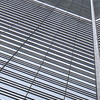}} &
			\subfloat[LR~\protect\linebreak(18.96/0.7246)]
			{\includegraphics[width = \patchSize, height = \patchSize]
				{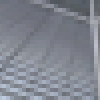}} &
			\subfloat[Bicubic~\protect\linebreak(19.21/0.7331)]
			{\includegraphics[width = \patchSize, height = \patchSize]
				{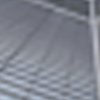}} &
			\subfloat[VDSR~\cite{vdsr_cvpr2016}~\protect\linebreak(19.94/0.7910)]
			{\includegraphics[width = \patchSize, height = \patchSize]
				{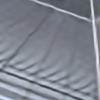}} \\ [-0.3cm]&
			\subfloat[MSLapSRN~\cite{lapsrn_pami2019}~\protect\linebreak(19.92/0.7894)]
			{\includegraphics[width = \patchSize, height = \patchSize]
				{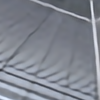}} &
			\subfloat[CARN~\cite{carn_eccv2018}~\protect\linebreak(20.82/0.8234)]
			{\includegraphics[width = \patchSize, height = \patchSize]
				{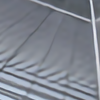}} &                   
			\subfloat[MSRN~\cite{msrn_eccv2018}~\protect\linebreak(21.11/0.8369)]
			{\includegraphics[width = \patchSize, height = \patchSize]
				{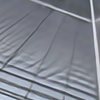}} &
			\subfloat[Ours~\protect\linebreak(\textbf{21.44/0.8447})]
			{\includegraphics[width = \patchSize, height = \patchSize]
				{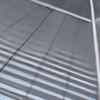}}
		\end{tabular}
		\vspace{-0.3cm}
		\begin{tabular}[b]{ccccc}
			\multirow{2}{*}[\rowArg]{
				\subfloat[image\_067 from Urban100]
				{\includegraphics[width = \fullSizeWid, height = \fullSizeHei]
					{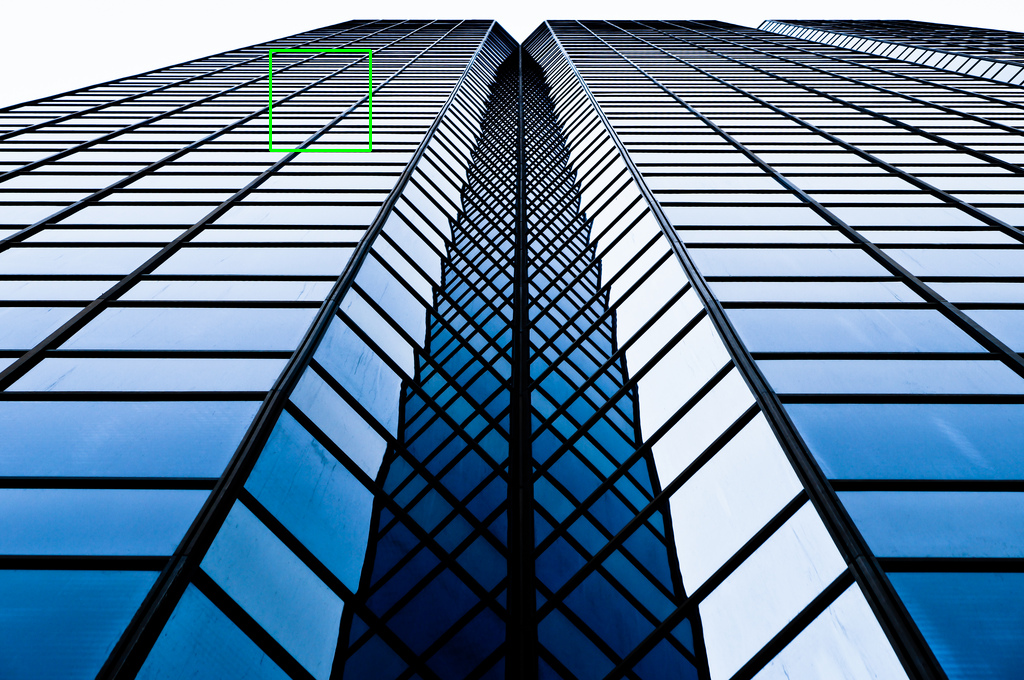}}} &
			\subfloat[HR~\protect\linebreak(PSNR/SSIM)]
			{\includegraphics[width = \patchSize, height = \patchSize]
				{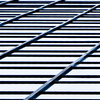}} &
			\subfloat[LR~\protect\linebreak(14.95/0.7116)]
			{\includegraphics[width = \patchSize, height = \patchSize]
				{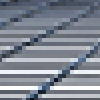}} &
			\subfloat[Bicubic~\protect\linebreak(15.80/0.7490)]
			{\includegraphics[width = \patchSize, height = \patchSize]
				{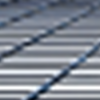}} &
			\subfloat[VDSR~\cite{vdsr_cvpr2016}~\protect\linebreak(17.30/0.8474)]
			{\includegraphics[width = \patchSize, height = \patchSize]
				{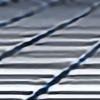}} \\ [-0.3cm]&
			\subfloat[MSLapSRN~\cite{lapsrn_pami2019}~\protect\linebreak(17.34/0.8577)]
			{\includegraphics[width = \patchSize, height = \patchSize]
				{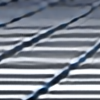}} &
			\subfloat[CARN~\cite{carn_eccv2018}~\protect\linebreak(18.12/0.8882)]
			{\includegraphics[width = \patchSize, height = \patchSize]
				{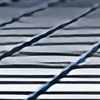}} &                   
			\subfloat[MSRN~\cite{msrn_eccv2018}~\protect\linebreak(18.58/0.8950)]
			{\includegraphics[width = \patchSize, height = \patchSize]
				{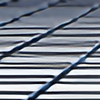}} &
			\subfloat[Ours~\protect\linebreak(\textbf{18.84/0.9035})]
			{\includegraphics[width = \patchSize, height = \patchSize]
				{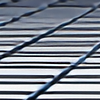}}
		\end{tabular}
		
		\begin{tabular}[b]{ccccc}
			\multirow{2}{*}[\rowArg]{
				\subfloat[image\_078 from Urban100]
				{\includegraphics[width = \fullSizeWid, height = \fullSizeHei]
					{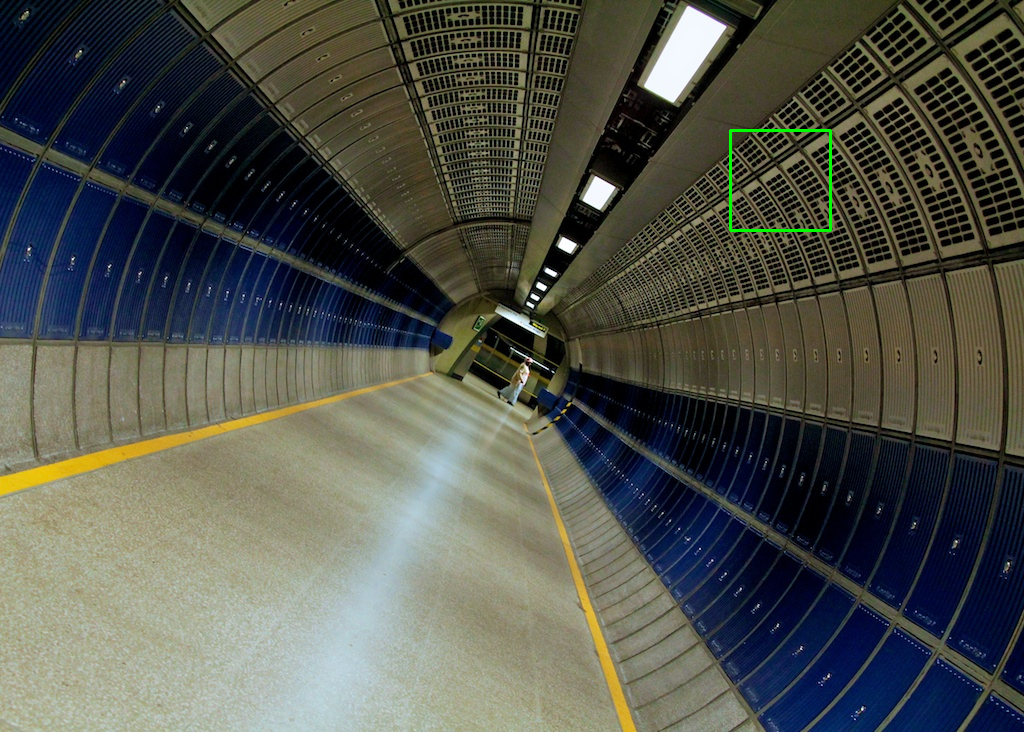}}} &
			\subfloat[HR~\protect\linebreak(PSNR/SSIM)]
			{\includegraphics[width = \patchSize, height = \patchSize]
				{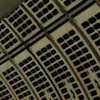}} &
			\subfloat[LR~\protect\linebreak(23.74/0.7624)]
			{\includegraphics[width = \patchSize, height = \patchSize]
				{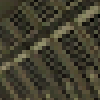}} &
			\subfloat[Bicubic~\protect\linebreak(24.49/0.7866)]
			{\includegraphics[width = \patchSize, height = \patchSize]
				{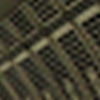}} &
			\subfloat[VDSR~\cite{vdsr_cvpr2016}~\protect\linebreak(25.49/0.8401)]
			{\includegraphics[width = \patchSize, height = \patchSize]
				{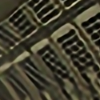}} \\ [-0.3cm]&
			\subfloat[MSLapSRN~\cite{lapsrn_pami2019}~\protect\linebreak(25.41/0.8395)]
			{\includegraphics[width = \patchSize, height = \patchSize]
				{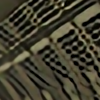}} &
			\subfloat[CARN~\cite{carn_eccv2018}~\protect\linebreak(25.88/0.8536)]
			{\includegraphics[width = \patchSize, height = \patchSize]
				{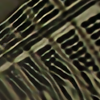}} &                   
			\subfloat[MSRN~\cite{msrn_eccv2018}~\protect\linebreak(26.12/0.8598)]
			{\includegraphics[width = \patchSize, height = \patchSize]
				{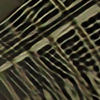}} &
			\subfloat[Ours~\protect\linebreak(\textbf{26.45/0.8658})]
			{\includegraphics[width = \patchSize, height = \patchSize]
				{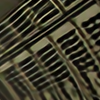}}
		\end{tabular}
	\end{center}
	\setlength{\abovecaptionskip}{0pt plus 2pt minus 2pt}
	\setlength{\belowcaptionskip}{0pt plus 2pt minus 2pt}
	\caption{Visualization comparisons on Urban100 with \textbf{BI}~$\times4$ degradation.}
	\label{fig:vis-result-bi}
\end{figure*}

Besides quantitative comparisons, we also analyze the qualitative restoration performance via visualization comparisons. Three images from Urban100 benchmark are chosen for comparison with \textbf{BI}~$\times4$ degradation, which is shown in Fig.~\ref{fig:vis-result-bi}. These images are from real world with abundant high-frequency textures and competitive for restoration with large scaling factors. From the result, SHSR can recover the structural information effectively, and find more accurate textures than other works.

\begin{figure*}[t]
	\captionsetup[subfloat]{labelformat=empty, justification=centering}
	\begin{center}
		\scriptsize
		\setlength\tabcolsep{0.1cm}
		\begin{tabular}[b]{cccc}
			\subfloat[HR~\protect\linebreak(PSNR/SSIM)]{\includegraphics[width=0.2\linewidth]{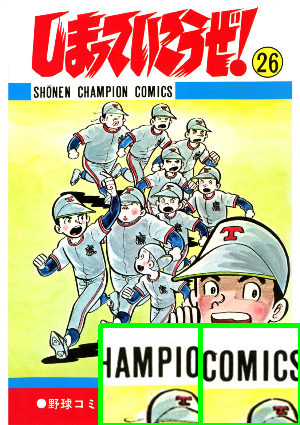}}&
			\subfloat[LR~\protect\linebreak(20.16/0.8521)]{\includegraphics[width=0.2\linewidth]{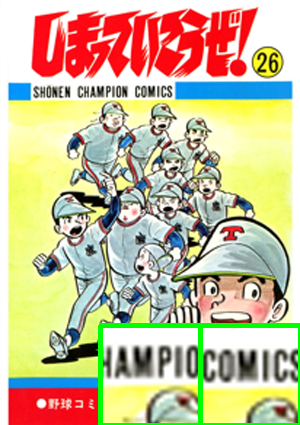}}&
			\subfloat[MSRN~\cite{msrn_eccv2018}~\protect\linebreak(26.31/0.9296)]{\includegraphics[width=0.2\linewidth]{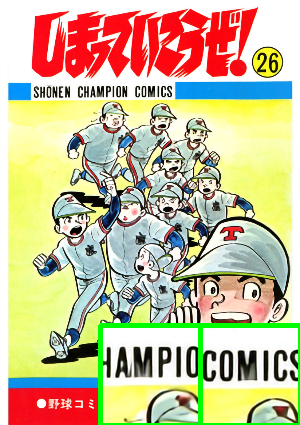}}&
			\subfloat[SHSR~\protect\linebreak(\textbf{26.46/0.9637})]{\includegraphics[width=0.2\linewidth]{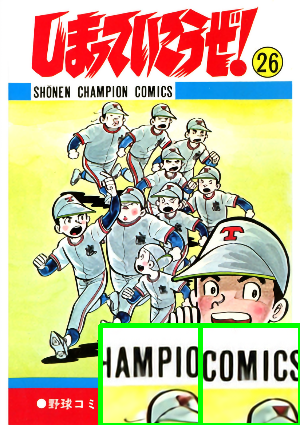}} \\ [-0.3cm]
			
			\subfloat[HR~\protect\linebreak(PSNR/SSIM)]{\includegraphics[width=0.2\linewidth]{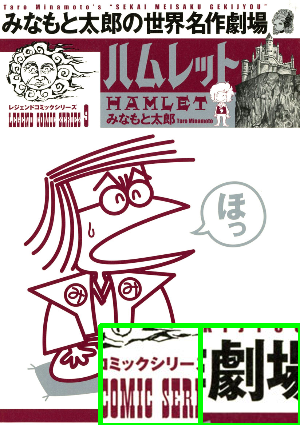}}&
			\subfloat[LR~\protect\linebreak(21.34/0.8693)]{\includegraphics[width=0.2\linewidth]{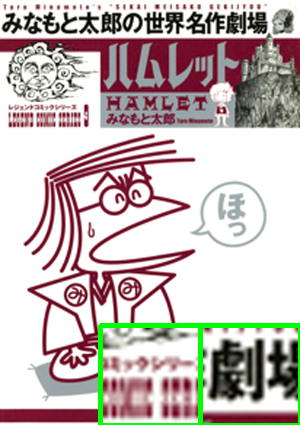}}&
			\subfloat[MSRN~\cite{msrn_eccv2018}~\protect\linebreak(27.49/0.9690)]{\includegraphics[width=0.2\linewidth]{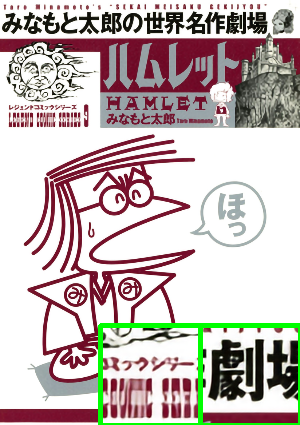}}&
			\subfloat[SHSR~\protect\linebreak(\textbf{27.76/0.9637})]{\includegraphics[width=0.2\linewidth]{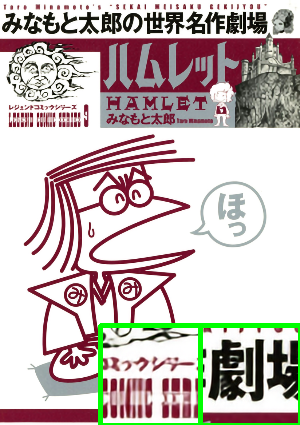}} \\
		\end{tabular}
	\end{center}
	\setlength{\abovecaptionskip}{0pt plus 2pt minus 2pt}
	\setlength{\belowcaptionskip}{0pt plus 2pt minus 2pt}
	\caption{Visualization comparisons on Manga109 with \textbf{BI}~$\times4$ degradation.}
	\label{fig:manga109}
\end{figure*}

Besides Urban100, we also conduct the experiments on Manga109, which is composed of comic book covers with plentiful line structures. The result is shown in Fig.~\ref{fig:manga109}. From the visualization comparison, SHSR recovers more lines and structural textures.

\begin{figure*}[t]
	\captionsetup[subfloat]{labelformat=empty, justification=centering}
	\begin{center}
		\scriptsize
		\setlength\tabcolsep{0.1cm}
		\begin{tabular}[b]{cccc}
			\subfloat[HR~\protect\linebreak(PSNR/SSIM)]{\includegraphics[width=0.2\linewidth]{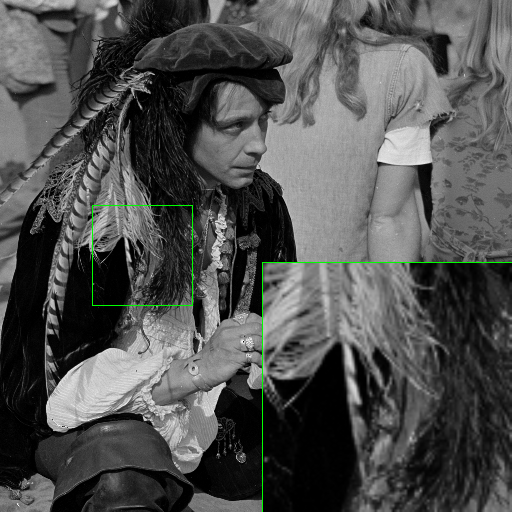}}&
			\subfloat[LR~\protect\linebreak(23.92/0.6380)]{\includegraphics[width=0.2\linewidth]{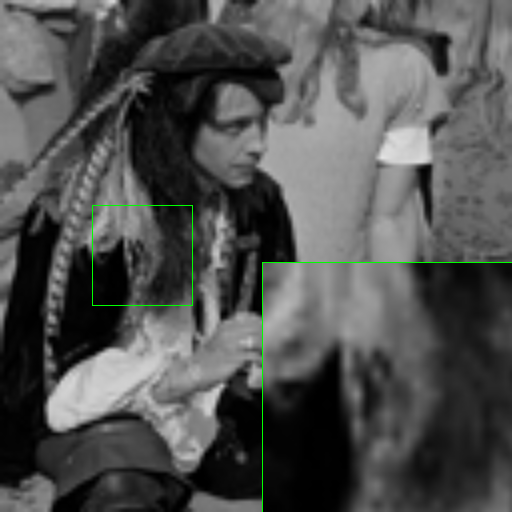}}&
			\subfloat[MSRN~\cite{msrn_eccv2018}~\protect\linebreak(26.11/0.7529)]{\includegraphics[width=0.2\linewidth]{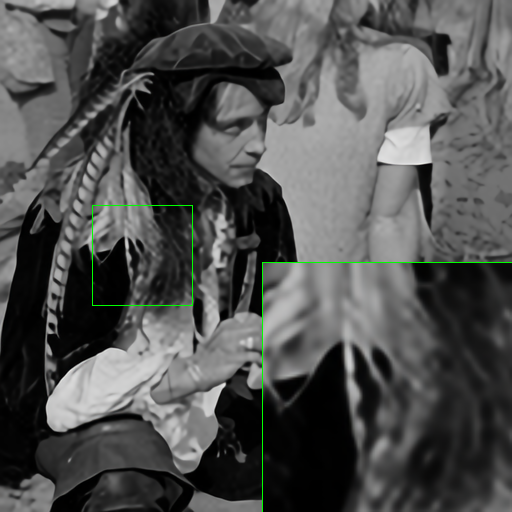}}&
			\subfloat[SHSR~\protect\linebreak(\textbf{26.18/0.7547})]{\includegraphics[width=0.2\linewidth]{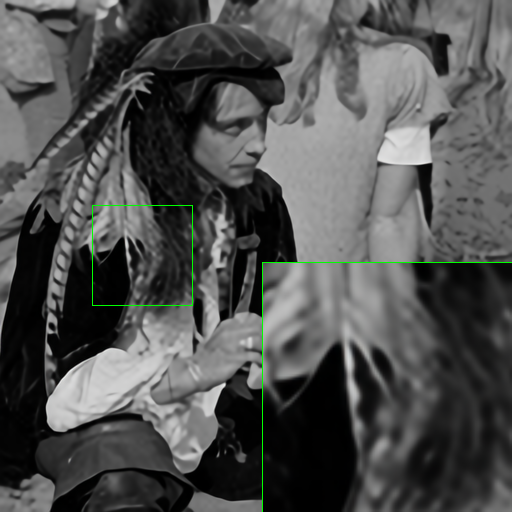}} \\
		\end{tabular}
	\end{center}
	\setlength{\abovecaptionskip}{0pt plus 2pt minus 2pt}
	\setlength{\belowcaptionskip}{0pt plus 2pt minus 2pt}
	\caption{Visualization comparisons on Set14 with \textbf{BI}~$\times4$ degradation.}
	\label{fig:set14}
\end{figure*}

Since Manga109 is a normal textured benchmark, we also compare the methods in the very textured situation, which is shown in Fig.~\ref{fig:set14}. The feather contains plentiful small lines and textures which are hard for recovery. From the comparison, SHSR can restore the textured image more accurately than MSRN.

\begin{figure*}[t]
	\captionsetup[subfloat]{labelformat=empty, justification=centering}
	\begin{center}
		\scriptsize
		\setlength\tabcolsep{0.1cm}
		\begin{tabular}[b]{ccccc}
			\subfloat[HR (PSNR)]    {\includegraphics[width=0.18\linewidth]{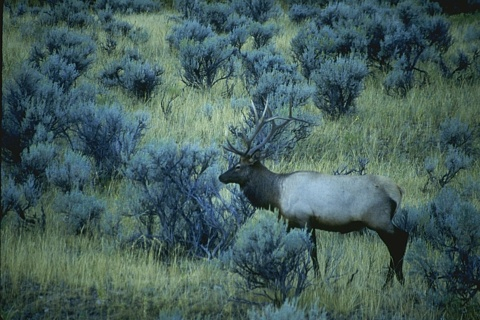}}&
			\subfloat[EDSR  (25.92)]{\includegraphics[width=0.18\linewidth]{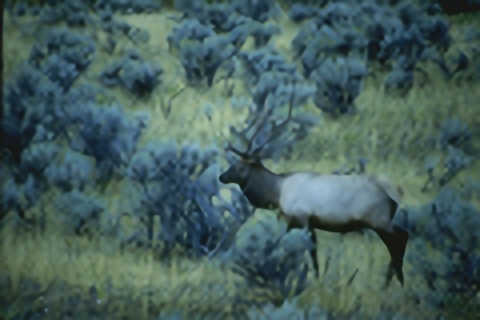}}&
			\subfloat[DBPN  (25.97)]{\includegraphics[width=0.18\linewidth]{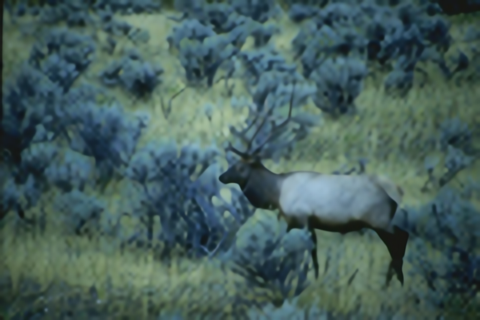}}&
			\subfloat[SRFBN (25.95)]{\includegraphics[width=0.18\linewidth]{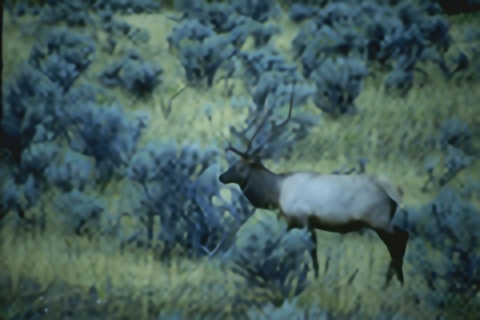}}&
			\subfloat[Ours  (25.94)]{\includegraphics[width=0.18\linewidth]{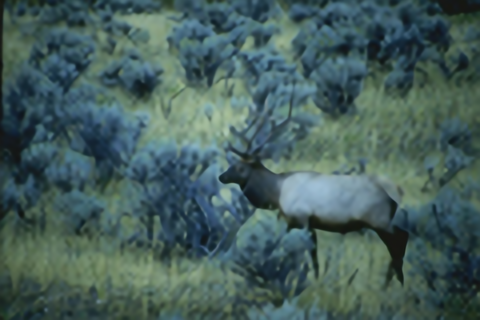}} \\ [-0.3cm]
			
			\subfloat[HR (PSNR)]    {\includegraphics[width=0.18\linewidth]{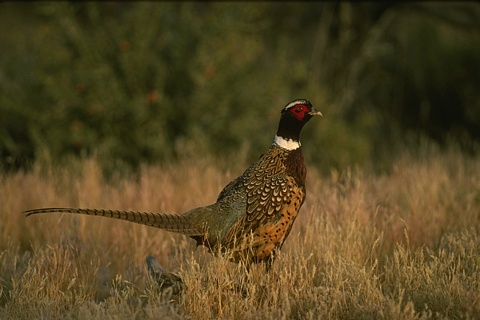}}&
			\subfloat[EDSR  (28.59)]{\includegraphics[width=0.18\linewidth]{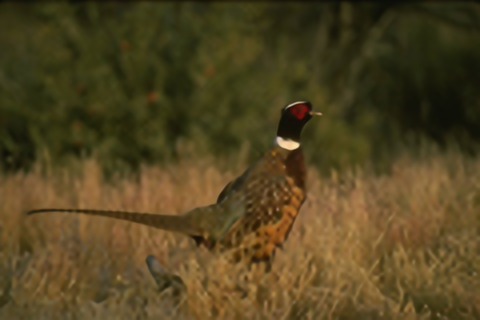}}&
			\subfloat[DBPN  (28.63)]{\includegraphics[width=0.18\linewidth]{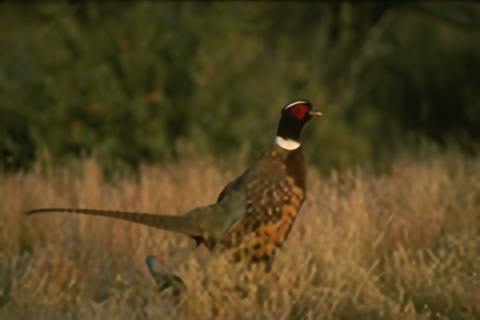}}&
			\subfloat[SRFBN (28.61)]{\includegraphics[width=0.18\linewidth]{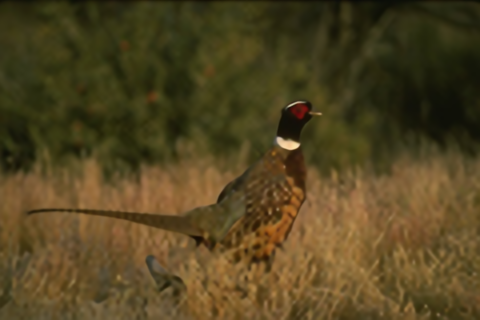}}&
			\subfloat[Ours  (28.60)]{\includegraphics[width=0.18\linewidth]{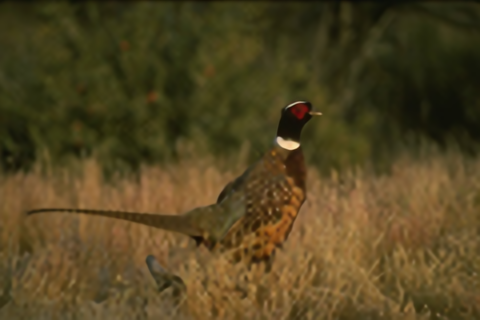}} \\ [-0.3cm]
			
			\subfloat[HR (PSNR)]    {\includegraphics[width=0.18\linewidth]{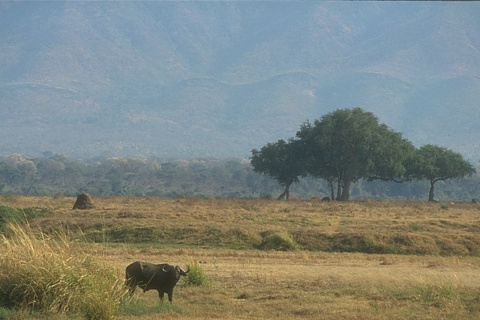}}&
			\subfloat[EDSR  (30.63)]{\includegraphics[width=0.18\linewidth]{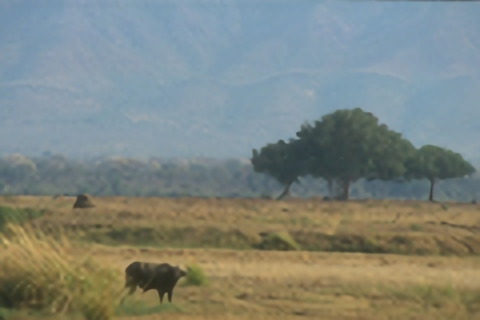}}&
			\subfloat[DBPN  (30.66)]{\includegraphics[width=0.18\linewidth]{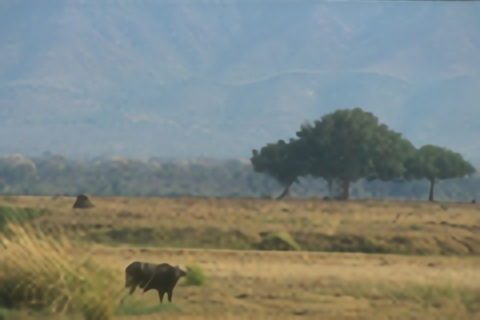}}&
			\subfloat[SRFBN (30.68)]{\includegraphics[width=0.18\linewidth]{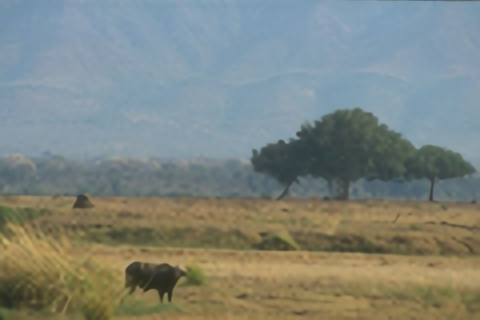}}&
			\subfloat[Ours  (30.69)]{\includegraphics[width=0.18\linewidth]{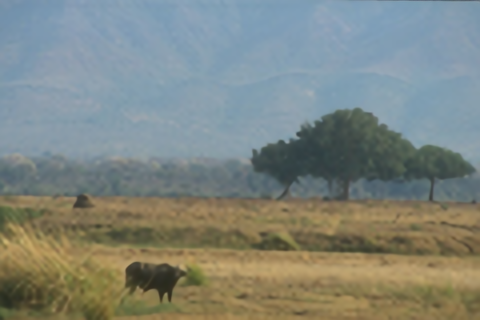}} \\ [-0.3cm]
			
			\subfloat[HR (PSNR)]    {\includegraphics[width=0.18\linewidth, height=0.12\linewidth]{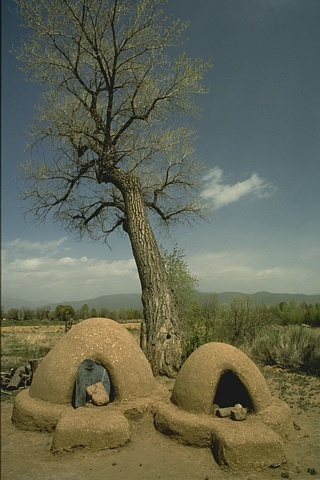}}&
			\subfloat[EDSR  (27.21)]{\includegraphics[width=0.18\linewidth, height=0.12\linewidth]{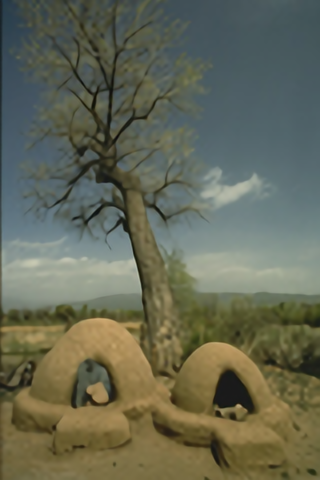}}&
			\subfloat[DBPN  (27.37)]{\includegraphics[width=0.18\linewidth, height=0.12\linewidth]{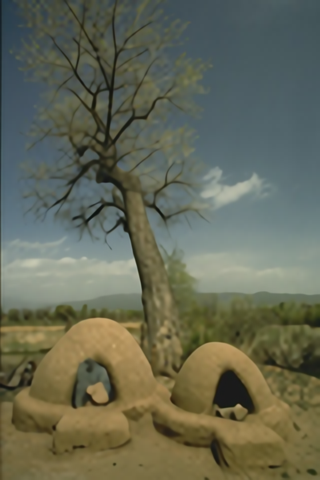}}&
			\subfloat[SRFBN (27.32)]{\includegraphics[width=0.18\linewidth, height=0.12\linewidth]{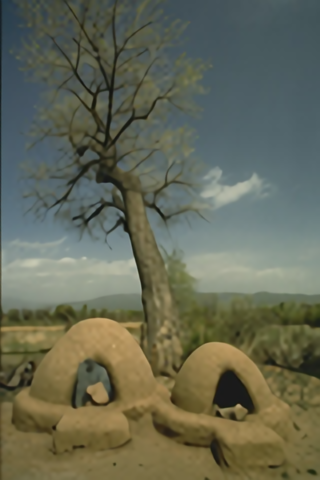}}&
			\subfloat[Ours  (27.30)]{\includegraphics[width=0.18\linewidth, height=0.12\linewidth]{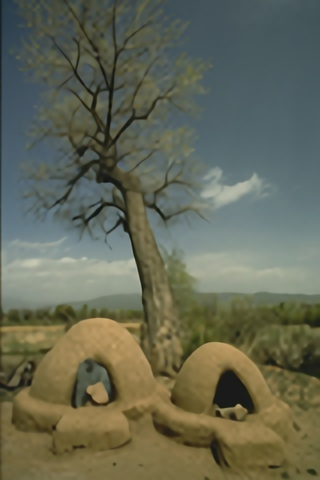}} \\
		\end{tabular}
	\end{center}
	\setlength{\abovecaptionskip}{0pt plus 2pt minus 2pt}
	\setlength{\belowcaptionskip}{0pt plus 2pt minus 2pt}
	\caption{Visualization comparisons on BSD100 benchmark with \textbf{BI}~$\times4$ degradation.}
	\label{fig:b100-r2}
\end{figure*}

To make the comparison more convincing, we compare the visualization performance on different objects, such as plant, grass and animals. Figure~\ref{fig:b100-r2} shows the visual results and PSNR results compared with EDSR, D-DBPN and SRFBN on BSD100 benchmark with various objects. It should be noted that EDSR, D-DBPN, and SRFBN are heavy networks and hold much more parameters and MACs than our network. Even though, our network can also achieve competitive or better visualization or objective performance than these works. 

Furthermore, we investigate the restoration capacity on blurred images. From the definition of SISR problem, these works can naturally handle the blurry issue which is regarded as a low-pass filter. We compare the proposed SHSR with recent works with blur and $\times3$ down-sampling~(\textbf{BD}$\times3$) degradation, which is shown in Tab.~\ref{tab:bd}. From the comparison, SHSR achieves competitive PSNR/SSIM performance with RDN, and the extension model SHSR$^+$ achieves superior performance than all other works. It should be noted that RDN holds 22,308K parameters and 2,282.2G MACs, which are much more than SHSR. From this perspective, SHSR is an efficient design which can effectively restore the blurred images.

\begin{table*}[t]
	\centering
	\caption{PSNR/SSIM results with \textbf{BD}$\times3$ degradation.}
	\label{tab:bd}
	\fontsize{6.5}{8}\selectfont
	\begin{tabular}{|c|c|c|c|c|c|c|c|c|}
		\hline
		\textbf{Method}& Bicubic& SRCNN~\cite{srcnn_pami2016}& VDSR~\cite{vdsr_cvpr2016}& IRCNN\_G~\cite{ircnn_cvpr2017}& IRCNN\_C~\cite{ircnn_cvpr2017}& RDN~\cite{rdn_pami2020}& SHSR& SHSR$^+$\\
		\hline
		\hline
		\textbf{Set5}& 			28.78/0.8308& 32.05/0.8944& 33.25/0.9150& 33.38/0.9182& 33.17/0.9157& 34.58/0.9280& 34.53/0.9274& \textbf{34.66/0.9284}\\
		\textbf{Set14}& 		26.38/0.7271& 28.80/0.8074& 29.46/0.8244& 29.63/0.8281& 29.55/0.8271& 30.53/0.8447& 30.51/0.8442& \textbf{30.60/0.8453}\\
		\textbf{B100}& 			26.33/0.6918& 28.13/0.7736& 28.57/0.7893& 28.65/0.7922& 28.49/0.7886& 29.23/0.8079& 29.22/0.8073& \textbf{29.28/0.8083}\\
		\textbf{Urban100}& 		23.52/0.6862& 25.70/0.7770& 26.61/0.8136& 26.77/0.8154& 26.47/0.8081& 28.46/0.8582& 28.48/0.8580& \textbf{28.63/0.8603}\\
		\textbf{Manga109}& 		25.46/0.8149& 29.47/0.8924& 31.06/0.9234& 31.15/0.9245& 31.13/0.9236& 33.97/0.9465& 34.05/0.9464& \textbf{34.36/0.9480}\\
		\hline
	\end{tabular}
\end{table*}

Besides the simulation degradation scenarios, we also compare the performance on real-world low-resolution samples without simulated sampling. Since there is no corresponding HR instance to calculate PSNR and SSIM, we choose three reference-free image quality evaluation indexes (NIQE~\cite{niqe_spl2013}, SSEQ~\cite{sseq_spic2014} and PI~\cite{esrgan_eccvw2018}) to estimate the image quality. We collect 10 low-resolution patches captured from the real-world, and build a benchmark to compare the reference-free image quality. To make the comparison more convincing, we compare our network with both the state-of-the-art bicubic-oriented SISR works (SRFBN~\cite{srfbn_cvpr2019} and USRNet~\cite{usrnet_cvpr2020}), and the real-world-oriented blind SR works (DAN~\cite{dan_nips2020}). The performances are shown in Table~\ref{tab:iqa}.

\begin{table*}[t]
	\centering
	\caption{Perceptual performance comparison on the real-world benchmark with scaling factor $\times4$. Lower value means better performance.}
	\label{tab:iqa}
	\fontsize{6.5}{8}\selectfont
	\begin{tabular}{|c||c|c||c||c|}
		\hline
		\textbf{Index}& \textbf{SRFBN~\cite{srfbn_cvpr2019}}& \textbf{USRNet~\cite{usrnet_cvpr2020}}& \textbf{DAN~\cite{dan_nips2020}} &\textbf{Ours}\\
		\hline
		\textbf{NIQE}& 10.09& 10.38& 11.49& 10.01 \\
		\textbf{SSEQ}& 57.90& 56.89& 48.01& 57.95 \\
		\textbf{PI}& 8.74& 8.87& 9.56& 8.70 \\
		\hline
		\textbf{MACs (G)}& 7746.1& 8545.8& -& 207.2 \\
		\textbf{Params (M)}& 3.63& 17.01& 4.32& 3.59 \\
		\hline
	\end{tabular}
\end{table*}

Compared with bicubic-oriented methods, our method achieves competitive or better perceptual performance with much fewer computation complexity. In the table, our network requires near 3\% MACs than SRFBN and USRNet and holds similar NIQE, SSEQ and PI performances. Compared with blind-SR methods, our network achieves similar perceptual performance to DAN with near 83\% parameters. The computation complexity of DAN is omitted because this method requires PCA operation before image restoration, which is difficult to calculate the MACs. It is worth noting that although DAN performs better than our method on SSEQ index, we are superior to DAN on NIQE and SSEQ indexes.

\section{Conclusion}
In this paper, we proposed a sequential hierarchical learning network~(SHSR) with fewer parameters and computation complexity for single image super-resolution~(SISR). Specially, a novel sequential multi-scale block~(SMB) was introduced in SHSR for hierarchical information exploration. Multi-scale combinations~(MCs) were designed in a recursive way according to the linearity of convolution operation, which progressively investigated the inter-scale correlation of hierarchical features. Besides the SMB, we also proposed a distribution transformation mechanism (DTB) to adjust the features and explored the correlations. Different from attention-based methods, scaling and bias factors were explored in parallel for better representations and performed in a normalization manner. Experimental results show SHSR could not only achieve competitive or better PSNR/SSIM results than other small works on five testing benchmarks, but also recover more complex structural textures. Meanwhile, the extension model SHSR$^+$ with much fewer parameters and lower computation complexity achieved competitive or better PSNR/SSIM results than other large networks.

\section*{Acknowledgement}
This work was supported in part by the National Natural Science Foundation of China under grant (62031013, 62072008); in part by the National Key Research and Development Project(2019YFF0302703); and in part by the High Performance Computing Platform of Peking University, which are gratefully acknowledged.

\bibliographystyle{ACM-Reference-Format}
\bibliography{main}


\end{document}